\documentclass[modern]{aastex631}

\usepackage{amsmath, amssymb, bm}
\usepackage{graphicx}
\usepackage[noabbrev]{cleveref}
\usepackage{appendix}

\DeclareMathOperator*{\argmin}{arg\,min}

\usepackage{array}
\newcolumntype{P}[1]{>{\centering\arraybackslash}p{#1}}

\DeclareMathOperator{\E}{\mathbb{E}}
\newcommand{\lr}[1]{\left( #1 \right)}
\newcommand{\tn}[1]{\textnormal{#1}}
\newcommand{\bh}[1]{\bm{\hat{#1}}}

\shortauthors{O'Leary et al.}

\graphicspath{{./}{figures/}}

\begin{document}

\title{Measuring the magnetic dipole moment and magnetospheric fluctuations of SXP 18.3 with a Kalman filter}

\author{Joseph O'Leary}
\affiliation{School of Physics, University of Melbourne, Parkville, VIC 3010, Australia.}
\affiliation{Australian Research Council Centre of Excellence for Gravitational Wave Discovery (OzGrav), Parkville, VIC 3010, Australia.}

\author{Andrew Melatos}
\affiliation{School of Physics, University of Melbourne, Parkville, VIC 3010, Australia.}
\affiliation{Australian Research Council Centre of Excellence for Gravitational Wave Discovery (OzGrav), Parkville, VIC 3010, Australia.}

\author{Nicholas J. O'Neill}
\affiliation{School of Physics, University of Melbourne, Parkville, VIC 3010, Australia.}
\affiliation{Australian Research Council Centre of Excellence for Gravitational Wave Discovery (OzGrav), Parkville, VIC 3010, Australia.}

\author{Patrick M. Meyers}
\affiliation{Theoretical Astrophysics Group, California Institute of Technology, Pasadena, CA 91125, USA.}

\author{Dimitris M. Christodoulou}
\affiliation{University of Massachusetts Lowell, Kennedy College of Sciences, Lowell, MA, 01854, USA.}
\affiliation{Lowell Centre for Space Science and Technology, Lowell, MA 01854, USA.}
\author{Sayantan Bhattacharya}
\affiliation{University of Massachusetts Lowell, Kennedy College of Sciences, Lowell, MA, 01854, USA.}
\affiliation{Lowell Centre for Space Science and Technology, Lowell, MA 01854, USA.}
\author{Silas G.T. Laycock}
\affiliation{University of Massachusetts Lowell, Kennedy College of Sciences, Lowell, MA, 01854, USA.}
\affiliation{Lowell Centre for Space Science and Technology, Lowell, MA 01854, USA.}

\begin{abstract}
The magnetic dipole moment $\mu$ of an accretion-powered pulsar in magnetocentrifugal equilibrium cannot be inferred uniquely from time-averaged pulse period and aperiodic X-ray flux data, because the radiative efficiency $\eta_0$ of the accretion is unknown, as are the mass, radius, and distance of the star. The degeneracy associated with the radiative efficiency is circumvented, if fluctuations of the pulse period and aperiodic X-ray flux are tracked with a Kalman filter, whereupon $\mu$ can be measured uniquely up to the uncertainties in the mass, radius, and distance. Here the Kalman filter analysis is demonstrated successfully in practice for the first time on {\em Rossi X-ray Timing Explorer} observations of the X-ray transient SXP 18.3 in the Small Magellanic Cloud, which is monitored regularly. The analysis yields $\mu = 8.0^{+1.3}_{-1.2} \, \times \, 10^{30} \, {\rm G \, cm^3}$ and $\eta_0 = 0.04^{+0.02}_{-0.01}$, compared to $\mu = 5.0^{+1.0}_{-1.0} \times 10^{30} \, {\rm G \, cm^3}$ as inferred traditionally from time-averaged data assuming $\eta_0=1$. The analysis also yields time-resolved estimates  of two hidden state variables, the mass accretion rate and the Maxwell stress at the disk-magnetosphere boundary. The success of the demonstration confirms that the Kalman filter analysis can be applied in the future to study the magnetic moments and disk-magnetosphere physics of  accretion-powered pulsar populations in the Small Magellanic Cloud and elsewhere.
\end{abstract}


\section{Introduction}\label{sec:intro}
The magnetic dipole moments $\mu$ of neutron stars span several orders of magnitude, from $\mu \sim 10^{33} \, {\rm G \, cm^3}$ for magnetars down to $\mu \sim 10^{26} \, {\rm G \, cm^3}$ for millisecond pulsars \citep{Lyne_2012}. In rotation-powered objects, $\mu$ is inferred from phase-connected pulse timing, assuming magnetic dipole braking \citep{Goldreich_1969,Ostriker_1969}. The resulting $\mu$ values are accurate up to a factor of order unity stemming from uncertainties in the electrodynamics of the plasma magnetosphere and the angle between the rotation and magnetic axes \citep{Spitkovsky_2006}, if the fundamental assumption of magnetic dipole braking does indeed hold. In accretion-powered objects, it is harder to measure $\mu$ accurately \citep{Mukherjee_2015}. Resonant electron cyclotron lines yield a direct measurement of the magnetic field strength in the line formation region near the stellar surface \citep{Makishima_2003,Caballero_2012,Revnivtsev_2016,Konar_2017,Staubert_2019} but are detected in relatively few objects. Otherwise one usually resorts to inferring $\mu$ from standard X-ray timing analysis and time-averaged statistics \citep{Ho_2014,Klus_2014,Mukherjee_2015}, coupled with predictions from physical theories of accretion; see Figure 6 in \cite{Klus_2014} and Figure 22 in \cite{Dangelo_2017} for example. In some objects, the local cyclotron measurements and global X-ray timing estimates of $\mu$ are performed simultaneously and calibrated successfully against each other; see Figure 11 in \cite{Makishima_1999}. Measurements of $\mu$ by the above methods form a key input into population synthesis calculations, whose aim is to present a consistent picture of how $\mu$ is distributed at birth and evolves, as a neutron star ages \citep{Faucher_2006,Kiel_2008}. \par
A popular approach for accretion-powered pulsars is to infer $\mu$ from time-averaged observations of the aperiodic X-ray flux $L\lr{t}$ and pulse period $P\lr{t}$, assuming magnetocentrifugal equilibrium \citep{Ho_2014,Klus_2014,Mukherjee_2015}. However the aforementioned approach relies on assuming a value for the radiative efficiency, which is unknown at the outset and  difficult to measure. Moreover, important time-dependent information is lost, when one analyses time-averaged data only, e.g.\ (anti)correlations like $\langle P\lr{t} L\lr{t'} \rangle$. Progress has occurred in certain directions, e.g.\ employing Monte Carlo simulations to estimate $\langle Y\lr{t} L\lr{t'} \rangle$, where  $Y\lr{t}$ is the measured pulsed fraction; see Figure 11 in \cite{Coe_2015} and Figure 2 in \cite{Yang_2018} for example. Further, there is evidence of a break in the X-ray flux autocorrelation function (and hence power spectral density) near the Kepler frequency at the disk-magnetosphere boundary for X-ray sources near spin equilibrium \citep{Revnivtsev_2009,Revnivtsev_2016}.  Interpreting all the above data is challenging because of uncertainties in the structure of the disk-magnetosphere boundary, which is complicated, time-dependent, and fundamentally three-dimensional, as revealed by modern numerical simulations \citep{Romanova_2003,Romanova_2015}.  In summary, time-averaged observables such as  $\langle L(t) \rangle$ and $\langle P(t) \rangle$ do not contain enough independent pieces of information to measure $\mu$ uniquely. \par 
In this paper, we generalize previous magnetocentrifugal estimates of $\mu$ for accretion-powered pulsars by exploiting the additional information available in the fluctuations of $L(t)$ and $P(t)$ around their time-averaged values, breaking the degeneracy between $\mu$ and the radiative efficiency. To do so, we apply the signal processing framework developed by \cite{Melatos_2022}, which (i) utilizes a Kalman filter to track fluctuations in three hidden state variables associated with magnetocentrifugal accretion (the mass accretion rate, the Maxwell stress at the disk-magnetosphere boundary and the radiative efficiency); and (ii) maximizes the Kalman filter likelihood to estimate the underlying, static physical parameters, including $\mu$. As this is the first time the method has been applied to astronomical data, we focus in this paper on demonstrating its feasibility using a single, regularly monitored object with sufficient samples of $L(t)$ and $P(t)$, namely the X-ray transient SXP 18.3 in the Small Magellanic Cloud (SMC) \citep{Corbet_2003}. The results presented herein differ from previous studies in the following two ways. (i) The degeneracy between $\mu$ and the radiative efficiency is broken, as noted above, so that $\mu$ and the radiative efficiency are measured independently. (ii) The Kalman filter framework exploits the auto- and cross-correlations between $P(t)$ and $L(t)$ to place multiple, simultaneous, interlocking constraints on the hidden state variables describing magnetocentrifugal accretion, breaking the degeneracies between them and producing a unique solution for their temporal evolution, corresponding to the most likely sequence of hidden states consistent with the time-ordered observational data. This approach preserves more statistical information than the alternative, which is to calculate $P(t)$ and $L(t)$ statistics (e.g. power spectral densities) separately and relate the associated ensemble averages to physical models of accretion \citep{Bildsten_1997,Riggio_2008,Klus_2014,Ho_2014,Mukherjee_2015,Serim_2023}. It builds on earlier work applying autoregressive moving average models to  torque-luminosity correlations \citep{Baykal_1993}, and testing for  consistency with random walk and shot noise processes \citep{deKool_1993,Baykal_1997,Lazzati_1997}. \par
The paper is organised as follows.  In Section \ref{sec:AD} we introduce the stochastic differential equations of motion which govern how the state variables describing accretion evolve in the context of the idealized, canonical, magnetocentrifugal picture \citep{Ghosh_1977}, as well as the measurement equations which map the state variables (some of which are hidden) to the measured aperiodic X-ray flux $L\lr{t}$ and pulse period $P\lr{t}$. The equations of motion and measurement equations are linearized about magnetocentrifugal equilibrium and cast into a state-space formulation suitable for Kalman filter analysis.  In Section \ref{sec:Example} we introduce the SMC X-ray transient SXP 18.3. We briefly describe its properties and the \textit{Rossi X-ray Timing Explorer} (RXTE) observations used throughout the analysis. We choose to focus our analysis on SXP 18.3 for two practical reasons: (i) the rotational state of the star has been classified by \cite{Yang_2017} as being approximately in equilibrium; and (ii) the RXTE Proportional Counter Array (PCA) sensor collected 854 observations over a period of $\sim 16$ years, comparable in volume to the representative test case involving synthetic data presented in \cite{Melatos_2022}. New estimates of the magnetic dipole moment $\mu$ and radiative efficiency $\eta_0$ associated with SXP 18.3 are presented in Section \ref{Sec:KFAnalysis}. Astrophysical implications are canvassed briefly in Section \ref{Sec:Conclusions} together with a note about future plans for generalizing the signal processing framework to population studies. \par 
\section{Measuring the magnetic moment} \label{sec:AD}
Disk accretion onto a magnetized, compact star is a complicated, time-dependent process involving hydromagnetic instabilities at the disk-magnetosphere boundary, which imprint a rich, three-dimensional, spatial structure on the system, as seen in magnetohydrodynamic numerical simulations \citep{Romanova_2003,Romanova_2005,Kulkarni_2008}.
The spatial structure in the simulations, e.g. the geometry of the magnetic field near the disk-magnetosphere boundary, cannot be inferred uniquely from the observables $P\lr{t}$ and $L\lr{t}$. In this paper, therefore, we choose to describe accretion in terms of the canonical, spatially averaged, magnetocentrifugal model \citep{Ghosh_1979}, motivated by the promising results presented in \cite{Melatos_2022}. The reader is directed to the latter reference for a full account of the model, its physical justification, and its implementation in the form of a Kalman filter; we do not repeat here the in-depth discussion in \cite{Melatos_2022}. Instead, in this section, we introduce briefly the variables, definitions, and dynamical equations needed to implement the Kalman filter, so that the interested reader is equipped to reproduce the key results of this paper, namely the independent measurements of $\mu$ and the radiative efficiency for SXP 18.3. In Section \ref{Sec:ObsStates}, we relate the observables $P\lr{t}$ and $L\lr{t}$ via nonlinear measurement equations to four time-dependent state variables, three of which are hidden. In Section \ref{Sec:Torque}, we discuss (i) the canonical magnetocentrifugal torque law; (ii) the degeneracy that exists between $\mu$ and the radiative efficiency for systems near magnetocentrifugal equilibrium; and (iii) a phenomenological, idealized model of the stochastic forces associated with instabilities at the disk-magnetosphere boundary, which drive the system away from equilibrium, first introduced in Section 2.4 in \cite{Melatos_2022}. The nonlinear measurement equations in Section \ref{Sec:ObsStates} and equations of motion in Section \ref{Sec:Torque} are linearized about the state of magnetocentrifugal equilibrium in Section \ref{Sec:LinearDynamics}, producing a state-space formulation of the problem which is suitable for Kalman filter analysis. The formulation in Section \ref{Sec:LinearDynamics} breaks the degeneracy between $\mu$ and the radiative efficiency. An explicit formula is presented to calculate $\mu$ as a function of the parameters estimated by the Kalman filter from the measurements $P(t)$ and $L(t)$ in Section \ref{sec:MagMomentObs}. A short review of the key elements of a linear Kalman filter is presented in Appendix \ref{sec:PE} for the convenience  of the reader. Details can be found in \cite{Melatos_2022}.
\subsection{Observables and state variables}\label{Sec:ObsStates}
In its simplest form, an X-ray timing experiment returns raw photon counts, which are barycenter corrected and converted to $N$ samples of the pulse period, $P\lr{t_{1}}, \hdots P\lr{t_{N}}$, and aperiodic X-ray luminosity, $L\lr{t_1},\hdots, L\lr{t_N}$, using standard X-ray timing analysis; see Section 2.3 in \cite{Yang_2017} for practical details about the data reduction pipeline. \footnote{In this paper, the distance $D = 62 \pm 0.3 \, \rm{kpc}$ to SXP 18.3 is known to an accuracy of $\approx 0.5\%$ \citep{Scowcroft_2016}. If $D$ is unknown, one works instead with the aperiodic X-ray flux $F_X\lr{t} = L\lr{t}/\lr{4 \pi D^2}$, rescaling the parameters to be estimated by powers of $D$ \citep{Melatos_2022}.} \par 
We express the standard magnetocentrifugal model of disk accretion \citep{Ghosh_1979} in terms of four time-dependent state variables. The angular velocity $\Omega\lr{t}$ of the neutron star is inversely related to the pulse period $P\lr{t}$. The other three state variables, defined as follows, are hidden. Let $Q\lr{t}$ be the rate at which matter flows from the accretion disk into the disk-magnetosphere boundary $\lr{\textnormal{units: } \textnormal{g} \; \textnormal{s}^{-1}}$. Let $S\lr{t}$ denote the Maxwell stress at the disk-magnetosphere boundary $\lr{\textnormal{units: } \textnormal{g} \; \textnormal{cm}^{-1} \; \textnormal{s}^{-2}}$. Let $\eta\lr{t}$ denote the dimensionless radiative efficiency with which the gravitational potential energy of material falling onto the star is converted into X-rays. \par 
Hidden state variables cannot be measured directly. Rather, they are related indirectly to the observables $P(t)$ and $L(t)$ through algebraic equations, which are nonlinear in general. The measured pulse period $P\lr{t}$ and aperiodic X-ray luminosity $L(t)$ are given respectively by
\begin{equation} \label{eq:SpinPeriod1}
    P\lr{t} = 2 \pi/\Omega\lr{t} + N_P \lr{t},
\end{equation}
and 
\begin{equation} \label{eq:Luminosity1}
    L\lr{t} = GM Q(t) \eta(t)/R  + N_L (t),
\end{equation}
where Newton's gravitational constant and the mass and radius of the neutron star are denoted by $G$, $M$, and $R$ respectively We assume that the additive noise terms $N_P\lr{t}$ and $N_L(t)$ are Gaussian and white, with $\langle N_P \lr{t_n} \rangle = 0$, $\langle N_L \lr{t_n} \rangle = 0$,  $\langle N_P \lr{t_n}N_P \lr{t_{n'}} \rangle = \Sigma_{PP}^2 \delta_{n,n'}$, $\langle N_L \lr{t_n}N_L \lr{t_{n'}} \rangle = \Sigma_{LL}^2 \delta_{n,n'}$, and $\langle N_P \lr{t_n}N_L \lr{t_{n'}} \rangle = 0$,  where $\delta_{n,n'}$ denotes the Kronecker delta. \par
The radiative efficiency $0 < \eta(t) < 1$ is the product of two factors: (i) the fraction of the specific gravitational potential energy $GM/R$ that is converted into X-rays in the observed band through radiation processes such as bremsstrahlung emission; and (ii) the fraction of $Q(t)$ which strikes the surface of the star, which does not equal unity in general, because some of $Q(t)$ is diverted into polar outflows (for example) by nonconservative hydromagnetic instabilities at the disk-magnetosphere boundary \citep{Matt_2005,Matt_2008,Romanova_2015}. It is a severe simplification to combine (i) and (ii) into a single scalar variable $\eta(t)$, as discussed in detail in Appendix C in \cite{Melatos_2022}. However, more realistic models \citep{Dangelo_2010,Dangelo_2012,Dangelo_2017} require more data than the $N= 854$ samples available for SXP 18.3. In the present analysis, therefore, we stick with $\eta\lr{t}$ and make the additional simplification $\eta\lr{t} = \eta_0 = {\rm constant}$, i.e.\ we hold $\eta\lr{t}$ constant in time but leave its value free to be estimated by the Kalman filter from the data. This approach is appropriate for the volume of data available. Despite its simplicity, it yields new astrophysical insights, namely the first independent measurement of $\eta_0$ and hence $\mu$ for SXP 18.3, as described in Section \ref{Sec:KFAnalysis}.
\subsection{Magnetocentrifugal accretion dynamics}\label{Sec:Torque}
The accretion torque acting on the star is governed by the location of the disk-magnetosphere boundary. In reality, the boundary has a complicated internal structure with nonzero thickness, which affects the torque and other dynamical variables [e.g.\ $\eta(t)$] in different ways \citep{Dangelo_2010,Dangelo_2012,Dangelo_2017}; see also Appendix C in \cite{Melatos_2022}. In this paper, guided by simplicity and the modest volume of data available, we adopt the canonical magnetocentrifugal model \citep{Ghosh_1977}, and approximate the boundary as having zero thickness and occurring at the Alfv\'{e}n radius $R_{\rm m}(t)$, where the Maxwell stress $S(t)$ balances the disk ram pressure, viz. $S\approx \rho v^2/2$, where $\rho = Q/\lr{4\pi R_{\rm m}^2 v}$ and $v = \lr{2 \, GM/R_{\rm m}}^{1/2}$ are the mass density and infall speed\footnote{The effective infall speed $v$ in different models and geometries differs by factors of order unity, e.g.\ \cite{Melatos_2022} take $v \approx (GM/R_{\rm m})^{1/2}$. \label{FN:EffectiveInfall}} respectively in free fall \citep{Menou_1999,Frank_2002}. The Alfv\'{e}n radius is given in terms of the state variables by
\begin{equation} \label{eq:AlfvenRad}
    R_{\rm m}\lr{t} = (2 \pi^{2/5})^{-1} \lr{GM}^{1/5} Q\lr{t}^{2/5} S\lr{t}^{-2/5}.
\end{equation}
The corotation radius,
\begin{equation} \label{eq:CoRot}
    R_{\rm c}\lr{t} = \lr{GM}^{1/3}\Omega\lr{t}^{-2/3},
\end{equation}
is located where Keplerian disk matter corotates with the neutron star. Reading Equation (\ref{eq:AlfvenRad}) at face value, it may seem that $R_{\rm m}(t)$ increases with $Q(t)$, which is the opposite of what one expects physically. However, $S(t)= (2\pi)^{-1} \mu^2 R_{\rm{m}}(t)^{-6}$ contains $R_{\rm{m}}(t)$ implicitly. Upon substituting the latter expression into Equation (\ref{eq:AlfvenRad}), we recover the standard expression (up to a factor of order unity) for the Alfv\'en radius in terms of the mass accretion rate and magnetic dipole moment; see for example Equation (7) in \cite{Klus_2014}. \par 
The sign of the magnetocentrifugal torque is governed by the ratio $R_{\rm m}/R_{\rm c}$, known as the fastness parameter. Material strikes the stellar surface for $R_{\rm m} < R_{\rm c}$, and the neutron star spins up in response to a combination of hydromagnetic and mechanical torques, with $\tn{d}{P}/\tn{d}t < 0$. The propeller mechanism occurs for $R_{\rm m} > R_{\rm c}$. Matter is ejected centrifugally by the corotating magnetosphere, causing the star to spin down, with $\tn{d}P / \tn{d}t > 0$. For $R_{\rm{m}} = R_{\rm{c}} = R_{\rm{m0}} = \, \rm{constant}$, the star is in magnetocentrifugal equilibrium, i.e.\ $Q\lr{t} = Q_0 = \rm constant$, $S\lr{t} = S_0 = \rm constant$, $\eta\lr{t} = \eta_0 = \rm constant$, with $\Omega(t) =  \Omega_0 = 2^{3/2} \pi^{3/5}\lr{GM}^{1/5}Q_0^{-3/5}S_0^{3/5}$ and $ L(t) = L_0 = GM Q_0 \eta_0 /R$. The three regimes $R_{\rm m}<R_{\rm c}$, $R_{\rm m}>R_{\rm c}$, and $R_{{\rm{m}}} = R_{{\rm{c}}}$ are described approximately  by the canonical magnetocentrifugal torque law \citep{Ghosh_1977} 
\begin{equation} \label{eq:spinequation1}
    I \frac{\tn{d} \Omega}{\tn{d}t} = \lr{GM}^{1/2} \left\{ 1 -  \left[\frac{R_{\rm m}(t)}{R_{\rm c}(t)}\right]^{3/2} \right\} R_{\rm m}(t)^{1/2}Q(t), 
\end{equation} 
where the star's moment of inertia is denoted by $I$. \Cref{eq:AlfvenRad,eq:CoRot,eq:spinequation1} feature the hidden variables $Q\lr{t}$ and $S\lr{t}$, but not $\eta\lr{t}$. For the sake of brevity, the reader is referred to (i) Section 2.2 in \cite{Melatos_2022} for a discussion about corrections to Equation  (\ref{eq:spinequation1}), e.g.\ due to radiation pressure \citep{Andersson_2005,Haskell_2015}, which are neglected in this paper, as the data are not plentiful enough to warrant their inclusion; and (ii) Appendices C.2 and C.3 in \cite{Melatos_2022}  for how to modify Equation (\ref{eq:spinequation1}) to incorporate important, time-dependent phenomena observed in accreting systems, e.g. due to the rapid onset of the propeller transition and disk trapping in the weak propeller regime \citep{Dangelo_2010,Dangelo_2012,Dangelo_2017}.   \par 
Assuming a magnetic dipole field, and imposing the zero-torque condition, $R_{{\rm{m}}}=R_{{\rm{c}}}=\, {\rm{constant}}$,  the star's magnetic moment $\mu$ is given by \footnote{The value of $\mu$ in Equation (\ref{eq:mu_equilibrium}) is $2^{-1/4}$ times its value in Equation (8) in \cite{Melatos_2022}, because we take the infall speed to be $v = (2 GM/R_{\rm m})^{1/2}$ in this paper and $v= ( GM/R_{\rm m})^{1/2}$ in \cite{Melatos_2022}. \label{FN:Mudifference}}
\begin{equation} \label{eq:mu_equilibrium}
    \mu = 2^{-3/4}\lr{GM}^{5/6}\Omega_0^{-7/6}Q_0^{1/2}.
\end{equation}
While we assume for simplicity that $\mu$ is constant, there is observational evidence to suggest that the magnetic dipole moments of recycled neutron stars decrease monotonically with accreted mass \citep{Shibazaki_1989}. Several mechanisms have been proposed to explain $\mu = \mu\lr{t} \neq \rm constant$, including accelerated ohmic decay \citep{Urpin_1995,Urpin_1997} and diamagnetic screening or burial \citep{uchida_1981,Hameury_1983,Brown_1998,Zhang_1998,Melatos_2001,Cumming_2001,Rai_2002,Payne_2004}. \par 
The zero-torque condition and the two time-averaged observables, $\langle 2\pi/P(t) \rangle = \Omega_0$ and $\langle L(t) \rangle = L_0$, do not contain enough independent pieces of information to uniquely solve for all four components of the state vector in equilibrium $\lr{\Omega_0, Q_0, S_0,\eta_0}$. Therefore one is left with two options: (i) assume a plausible value for one component, e.g.\ $\eta_0 = 1$; or (ii) exploit the time-dependent information in the observations $P(t)$ and $L(t)$, in addition to $\langle P(t) \rangle$ and $\langle L(t) \rangle$, and treat one state vector component as an unknown to be estimated from the data. There are at least two approaches available to utilize the time-dependent information in $P(t)$ and $L(t)$, namely  $\chi^2$ minimization \citep{Takagi_2016,Yatabe_2018} and Kalman filter analysis \citep{Melatos_2022}. Motivated by the results in \cite{Melatos_2022}, we adopt the latter technique, yielding the first independent estimate of $\eta_0$ and hence $\mu$ for SXP 18.3. In Appendix \ref{App:GhoshLambApp} we follow \cite{Takagi_2016} and \cite{Yatabe_2018} and also employ $\chi^2$ minimization to estimate $\mu$ independently for SXP 18.3 and compare the output of both techniques for completeness.   \par
Observations of accretion-powered pulsars reveal pronounced, stochastic variability in the measured aperiodic X-ray flux $L\lr{t}$ and pulse period $P\lr{t}$ \citep{Bildsten_1997}. The variability is driven by several mechanisms, discussed in detail in Section 2.4 in \cite{Melatos_2022}.  Here we focus on  mean-reverting fluctuations about magnetocentrifugal equilibrium, which are typical of self-limiting instabilities and are observed routinely in the output of three-dimensional magnetohydrodynamic simulations \citep{Romanova_2005,Romanova_2021}.  Accordingly, we adopt an idealized, phenomenological, Ornstein-Uhlenbeck model \citep{Gardiner_1985} for the state variables $Q(t)$ and $S(t)$, and assume that they satisfy the Langevin equations \citep{Melatos_2022}
\begin{equation}
    \frac{\tn{d}Q}{\tn{d}t} = -\gamma_Q [Q(t) - Q_0] + \xi_Q\lr{t}, \label{eq:LE_Q}
\end{equation}
\begin{equation}
    \frac{\tn{d}S}{\tn{d}t} = -\gamma_S [S(t) - S_0] + \xi_S\lr{t}. \label{eq:LE_S}
\end{equation}
In Equations (\ref{eq:LE_Q}) and (\ref{eq:LE_S}), $\gamma_Q$ and $\gamma_S$ are damping constants, and $\xi_Q(t)$ and $\xi_S(t)$ are fluctuating driving terms satisfying $\langle \xi_Q(t) \rangle = 0$, $\langle \xi_S(t) \rangle = 0$, $\langle \xi_Q(t) \, \xi_Q(t') \rangle = \sigma_{QQ}^2 \, \delta(t-t')$, $\langle \xi_S(t) \, \xi_S(t') \rangle = \sigma_{SS}^2 \, \delta(t-t')$, and $\langle \xi_Q(t) \, \xi_S(t') \rangle = 0$.  Equations (\ref{eq:LE_Q}) and (\ref{eq:LE_S}) are supplemented with the deterministic equation
\begin{equation} \label{eq:simple_eta}
    \eta \lr{t} = \eta_0,
\end{equation}
where the reader is reminded that $\eta_0$ is constant but free to be estimated with the Kalman filter in Section \ref{Sec:KFAnalysis}. The Langevin equations (\ref{eq:LE_Q}) and (\ref{eq:LE_S}) ensure that $Q\lr{t}$ and $S\lr{t}$ wander randomly about their equilibrium values, and do not undergo long-term secular drifts,  with characteristic timescales of mean reversion given by  $\gamma_Q^{\,-1}$ and $\gamma_S^{\,-1}$, and rms fluctuations $\sim \gamma_Q^{\,-1/2}\sigma_{QQ}$ and $\gamma_S^{\,-1/2}\sigma_{SS}$ respectively. The statistics associated with  $\xi_{Q}\lr{t}$ and $\xi_{S}\lr{t}$  are discussed in more detail in Appendix \ref{sec:PE}. \par 
Equations (\ref{eq:LE_Q}) and (\ref{eq:LE_S}) are highly idealized. They omit some important pieces of microphysics operating at the disk-magnetosphere boundary \citep{Romanova_2015}, which are discussed in detail in Section 2.4 and Appendix C of \cite{Melatos_2022}.  They compound the idealizations inherent in the torque law (\ref{eq:spinequation1}), discussed above, and in Equation (\ref{eq:simple_eta}) through the definition and constancy of $\eta(t)$ (see Section \ref{Sec:ObsStates}). Altogether, Equations (\ref{eq:spinequation1}) and (\ref{eq:LE_Q})--(\ref{eq:simple_eta}) represent a compromise between realism and the explanatory power of the available volume of data, specifically \ $N = 854$ samples for SXP 18.3. 
\subsection{Linear state-space formulation}\label{Sec:LinearDynamics}
The Kalman filter is a standard tool for inferring the most probable trajectory of a set of stochastic hidden variables, such as $\Omega(t_n)$, $Q(t_n)$, and $S(t_n)$, given a data stream of noisy measurements, such as $P(t_n)$ and $L(t_n)$, a set of measurement equations, such as (\ref{eq:SpinPeriod1}) and (\ref{eq:Luminosity1}), and a set of stochastic evolution equations, such as (\ref{eq:spinequation1}), (\ref{eq:LE_Q}), and (\ref{eq:LE_S}) \citep{Gelb_1974,Wan_2000}. The key elements of the linear Kalman filter are reviewed in abridged form in Appendix \ref{sec:PE} for the convenience of the reader; details specific to this application can be found in \cite{Melatos_2022}. In this paper, the measurement equations (\ref{eq:SpinPeriod1}) and (\ref{eq:Luminosity1}) and the magnetocentrifugal torque law (\ref{eq:spinequation1}) are nonlinear. Hence one must choose between two approaches: (i) apply a nonlinear estimation algorithm, e.g.\ a particle filter \citep{Carpenter_1999, Gustafsson_2010}, or an extended or unscented Kalman filter; see Sections 2 and 3 in \cite{Wan_2000} and Section 2 of \cite{Kandepu_2008}; or (ii) linearize Equations (\ref{eq:SpinPeriod1}), (\ref{eq:Luminosity1}), and (\ref{eq:spinequation1}) about magnetocentrifugal equilibrium, under the assumption that fluctuations are small, i.e.\ $\gamma_A^{-1/2}\sigma_{AA} \ll A_0$, with $A \in \{Q,S\}$. As this paper reports the first application of the Kalman filter framework to astronomical data, and aims to supply a proof of principle, we adopt approach (ii) to be conservative and defer approach (i) to future work. Nonlinear analyses are straightforward in principle but are more expensive computationally; see Appendices B and C in \cite{Melatos_2022} for illustrative examples applied to synthetic data. \par 
Linearizing the equations of motion (\ref{eq:spinequation1}), (\ref{eq:LE_Q}), and (\ref{eq:LE_S}), and introducing the notation $A_1 = \left[A\lr{t} - A_0 \right]/A_0$ with $A \in \{P, L, \Omega, Q, S \}$ for perturbed quantities, we obtain \citep{Melatos_2022}
\begin{equation} \label{eq:EoMTrunc}
    \frac{\tn{d}}{\tn{d}t}
    \begin{pmatrix}
    \Omega_1 \\ Q_1 \\ S_1
    \end{pmatrix}
    =
    \begin{pmatrix}
    -\gamma_\Omega & -3\gamma_\Omega/5 & 3\gamma_\Omega/5 \\
    0 & -\gamma_Q & 0 \\
    0 & 0 & -\gamma_S
    \end{pmatrix}
    \begin{pmatrix}
    \Omega_1 \\ Q_1 \\ S_1
    \end{pmatrix}
    + 
    \begin{pmatrix}
    0 \\ Q_0^{-1}\xi_Q \\ S_0^{-1}\xi_S
    \end{pmatrix},
\end{equation}
with 
\begin{equation}\label{eq:GammaOmegaFull}
    \gamma_\Omega = \frac{(GM\, R_{\rm m0})^{1/2} Q_0}{I\Omega_0}.
\end{equation}
Linearizing the measurement equations (\ref{eq:SpinPeriod1}) and (\ref{eq:Luminosity1}), we obtain
\begin{equation} \label{eq:ME_TruncP}
    P_1 = -\Omega_1 + P_0^{-1}N_P,
\end{equation}
and 
\begin{equation} \label{eq:ME_TruncL}
    L_1 = Q_1 + L_0^{-1}N_L.
\end{equation}
Equations (\ref{eq:EoMTrunc}), (\ref{eq:ME_TruncP}), and (\ref{eq:ME_TruncL}) now fit readily into the linear Kalman filter framework discussed in Appendix \ref{sec:PE}. They are a simplified version of what appears in \cite{Melatos_2022}, in that we set $\eta_1(t) = [\eta(t) - \eta_0]/\eta_0 = 0$ and $\gamma_\eta \rightarrow \infty$ in Equations (13), (16), (19), and (D11) in \cite{Melatos_2022} in order to keep $\eta(t)$ constant in time but free to be estimated. The simplifications are justified by the limited sample sizes available with modern X-ray timing experiments, i.e.\ $854$ samples for SXP 18.3. \par 
Although we elect to track $\bm{X}(t) = [\Omega(t), Q(t), S(t), \eta(t)]$ with a linear Kalman filter in the present paper, the specific choice of state-space variables is not unique. For example, one may choose to replace Equation (\ref{eq:LE_S}) with an Ornstein-Uhlenbeck process for the Alfv\'en radius, postulating mean-reverting stochastic fluctuations in $R_{\rm{m}}(t)$ instead of $S(t)$. In practice, it is unclear from theory or observations what approach better approximates the accretion dynamics in real systems. To support future Kalman filter applications using alternative state-space variables, a linear state-space formulation using $R_{\rm{m}}(t)$ in lieu of $S(t)$ is presented in Appendix \ref{App:AlternativeSSVars} for the benefit of the reader.
\subsection{Magnetic dipole moment from the Kalman filter output}\label{sec:MagMomentObs}
 The five fundamental model parameters estimated by the Kalman filter from the data are $\mathbf{\Theta} = \lr{\gamma_\Omega,\gamma_A, \sigma_{AA}}$, with $A \in \left\{Q, S \right\}$. One of these, namely $\gamma_\Omega$, is combined with the sample means $\Omega_0 = 2^{3/2} \pi^{3/5}\lr{GM}^{1/5}Q_0^{-3/5}S_0^{3/5}$ and $L_0 = GM Q_0 \eta_0 /R$,
  and the magnetocentrifugal equilibrium condition, to uniquely solve for  $\lr{\Omega_0,Q_0,S_0,\eta_0}$, adopting approach (ii) in Section \ref{Sec:Torque}. The explicit expressions for $Q_0$, $S_0$, and $\eta_0$ follow from the step-by-step guide outlined in Appendix A in \cite{Melatos_2022} with one minor modification. The Maxwell stress component,
  \begin{equation}\label{Eq:S0_denominator}
  S_0 = \frac{I \Omega_0^3 \gamma_{\Omega}}{2^{5/2} \pi GM}, 
  \end{equation}
  used in this paper is $2^{-1/2}$ times its value in Equation (A5) in Appendix A in \cite{Melatos_2022}; see Equation (\ref{eq:AlfvenRad}) and Footnotes \ref{FN:EffectiveInfall} and \ref{FN:Mudifference} for details. The expressions for $Q_0$ and $\eta_0$ remain unchanged and are not quoted here for the sake of brevity;  see Equations (A4) and (A6) in \cite{Melatos_2022}. The magnetic moment $\mu$ then follows from Equation (\ref{eq:mu_equilibrium}). 
\section{SXP 18.3: a demonstration} \label{sec:Example}
The SMC is a dwarf irregular galaxy orbiting the Milky Way at a known distance of $ 62 \; \pm \; 0.3 \; \textnormal{kpc}$ \citep{Scowcroft_2016}. It hosts more than $120$ known high-mass X-ray binaries (HMXBs) \citep{Haberl_2022}; see \cite{White_1995} and \cite{Reig_2011} for overviews of X-ray binaries. The optical counterpart is usually a Be star, with only one system hosting a supergiant donor star, namely  SMC X-1 \citep{Haberl_2016}.  The large population of SMC Be-type HMXBs is ascribed to: (i) recent star formation \citep{Galache_2008}; (ii) low metallicity  \citep{Antoniou_2010}; and (iii) recent tidal interactions with the Large Magellanic Cloud \citep{Gardiner_1996}. The orbital parameters and stellar properties of SMC HMXBs have been studied extensively \citep{Galache_2008,Townsend_2011,Klus_2014,Coe_2015,Christodoulou_2016, Christodoulou_2017} with space-based X-ray observatories, e.g.\ the RXTE PCA \citep{Jahoda_2006}.  In Section \ref{sec:SourceProperties} we discuss the population of SMC HMXBs analyzed by \cite{Yang_2017}, and introduce the X-ray transient SXP 18.3. Details of the RXTE PCA measurements of SXP 18.3 analyzed with the Kalman filter are presented in Section \ref{sec:RXTEObs}. 
\subsection{Source properties}\label{sec:SourceProperties}
Archival X-ray time-of-arrival measurements collected between 1997 and 2014 by the RXTE PCA, \textit{XMM-Newton} European Photon Imaging Camera \citep{Struder_2001}, and \textit{Chandra} Advanced CCD Imaging Spectrometer \citep{Garmire_2003} are publicly available via the High Energy Astrophysics Science Archive Research Centre.\footnote{\url{https://heasarc.gsfc.nasa.gov/}}
They are also post-processed into pulse period, pulse amplitude, pulsed fraction, and aperiodic X-ray luminosity time series by \cite{Yang_2017}. The latter authors estimate the pulse period derivative $\tn{d}P/\tn{d}t$ from a linear fit of $P(t_n)$ versus $t_n$ for $53$ stars with enough data. They characterise the torque distribution of the population in terms of $\tn{d}P/\tn{d}t$ divided by its measurement error, denoted by $\epsilon$ in Table 3 in \cite{Yang_2017}. They classify $12$ pulsars as spinning up $\lr{\epsilon \leq -1.5}$, $11$ as spinning down $\lr{\epsilon \geq 1.5}$, and $30$ as being near spin equilibrium $\lr{-1.5 < \epsilon < 1.5}$,  labelled by the letters ``U'', ``D'', and ``C'' respectively in Table 3 in \cite{Yang_2017}. \par
Using the post-processed time series  in \cite{Yang_2017}, we select a subpopulation of accretion-powered pulsars which satisfy two conditions. (i) The number of RXTE PCA samples is $\geq 850$, which is sufficient for the Kalman filter to converge accurately, as demonstrated by \cite{Melatos_2022}; see Table 2 in \cite{Yang_2017}. (ii) There is no  long-term secular drift in the spin period, i.e.\ each source is near equilibrium; see Table 3 in \cite{Yang_2017}. Although the Kalman filter framework can be applied to accretion-powered pulsars in magnetocentrifugal disequilibrium, as confirmed for synthetic data in Appendix B of \cite{Melatos_2022}, we elect to focus on near-equilibrium systems in this paper to be conservative, as this is the first time the framework is applied to real astronomical data.  Upon imposing criteria (i) and (ii), we are left with a
subsample of nine objects, whose names and timing properties are listed in Table \ref{Table:SourceProperties}. \par
We focus our analysis on SXP 18.3 for three practical reasons. (i) It has the largest number of significant detections, $N_{\rm det} = 73$, in Table \ref{Table:SourceProperties}. The concept of significance is defined below. (ii) The standard deviation of the pulse period satisfies $\sigma_P = 0.025 \, {\rm s} \ll P_0$ (see Table \ref{Table:SourceProperties}), consistent with the assumption of linearity. (iii) It accretes via a disk, consistent with the canonical picture of magnetocentrifugal accretion. With respect to point (iii) above, \cite{Klus_2014} investigated whether the X-ray emission is powered by an accretion disk or stellar wind \citep{Shakura_2012} for 42 SMC HMXBs, including SXP 18.3. They estimated the relative velocity $V_{\rm{rel}}$ between the accreted matter, calculated from the stellar wind velocity at the companion star's circumstellar disk, and the Keplerian orbital velocity of the neutron star. They compared $V_{\rm{rel}}$  with the critical relative velocity $V_{\rm{Crel}}$, below which disk accretion occurs (i.e.\ $R_{\rm{m}} < R_{\rm{c}}$), and found $V_{\rm{rel}}/V_{\rm{Crel}} < 1$ for the 42 HMXBs, indicating that all 42 systems accrete via a disk instead of a wind; see Section 3 and Figure 5 in \cite{Klus_2014} for further details. With respect to point (i) above, the statistical significance of each timing point is determined by searching the light curve for pulsations and calculating the number of independent frequencies and the Lomb-Scargle power according to Equation (2) in \cite{Yang_2017}; see Section 2.4.1 and Table 3 in \cite{Yang_2017}, as well as \cite{VanderPlas_2018}  for how to detect periodic signals with Lomb-Scargle periodograms. In X-ray timing analysis, it is standard practice to preference observations with statistical significance $\geq 99\%$.   \par
SXP 18.3 was discovered by \cite{Corbet_2003}, who detected pulsations with periods $18.37 \pm 0.01 \; \textnormal{s}, 18.36 \pm 0.01 \; \textnormal{s}$ and $18.37 \pm 0.1 \; \textnormal{s}$ on $2003$ November $26$, $2003$ December $4$, and $2003$ December $11$ respectively, using  the RXTE PCA. SXP 18.3 is categorised in Table 3 in \cite{Yang_2017} as being near magnetocentrifugal equilibrium;  a linear fit of  $P(t_n)$ versus $t_n$ yields $\tn{d}P/\tn{d}t \approx \left(3 \pm 4\right) \times 10^{-6} \; \textnormal{s}\; \textnormal{day}^{-1}$ and $\epsilon = 0.75$. \cite{Klus_2014} estimated the surface magnetic field strength $B = 5.0^{+1.0}_{-1.0} \,  \times 10^{12} \, \textnormal{G}$ of SXP 18.3  and hence inferred $\mu = 5.0^{+1.0}_{-1.0} \times  10^{30} \; \rm{G} \; \rm{cm}^3$ based on $Q_0 \propto \langle L(t) \rangle$ and $\eta_0 = 1$; see Section 4 and Table 3 in \cite{Klus_2014} for further details. 
\begin{table}[t]
    \hspace{1.0cm}
    \begin{tabular}{cccccc}
    \hline
    Pulsar & $\dot{P}$ ($\rm{s \; day^{-1}}$) & $\epsilon$ & $\sigma_P$ ($\rm{s}$) & $N$ & $N_{\rm det}$ \\
    \hline
    SXP 5.05 & $-6.0\times 10^{-6}$ & $-0.50$ & $0.021$ & $889$ & $5$ \\
    SXP 18.3 & $3.0\times 10^{-6}$ & $0.75$ & $0.025$ & $854$ & $73$ \\
    SXP 74.7 & $7.7\times 10^{-5}$ & $0.57$ & $0.97$ & $854$ & $21$ \\
    SXP 82.4 & $5.3\times 10^{-5}$ & $0.65$ & $0.28$ & $863$ & $19$ \\
    SXP 140 & $-1.1\times 10^{-4}$ & $-0.21$ & $0.84$ & $879$ & $7$ \\
    SXP 214 & $-8.9\times 10^{-5}$ & $-0.45$ & $1.1$ & $854$ & $16$ \\
    SXP 293 & $-5.0\times 10^{-5}$ & $-0.29$ & $0.77$ & $944$ & $13$ \\
    SXP 504 & $9.0\times 10^{-6}$ & $0.02$ & $3.7$ & $909$ & $32$ \\
    SXP 565 & $2.3\times 10^{-4}$ & $0.20$ & $4.5$ & $872$ & $12$ \\
    \hline
    \end{tabular}
	\caption{Candidate SMC HMXBs for Kalman filter analysis. The pulse period derivative (column $2$), proximity to spin equilibrium (column $3$), pulse period standard deviation (column $4)$, total number of RXTE PCA observations (column $5$), and number of significant detections (column 6) are denoted by $\dot{P}$, $\epsilon$, $\sigma_P$, $N$, and $N_{\rm det}$ respectively. The subsample in Table \ref{Table:SourceProperties} is defined by $|\epsilon| \leq 1.5$ and $N \geq 850$.}
	\label{Table:SourceProperties}
\end{table}
\subsection{RXTE data}\label{sec:RXTEObs}
The RXTE PCA is a non-imaging detector \citep{Corbett_2008}. The recorded photon counts are converted to light curves using standard X-ray timing analysis; see \cite{VanderKlis_1989} for further details, and Sections 1.3 and 4.4.1 in \cite{Laycock_2005} and \cite{Patruno_2021} respectively. Once extracted, the light curves are referred to the Solar System barycenter with background noise subtracted \citep{Yang_2017}. To estimate the spin period, the light curves are searched for significant pulsations using the Lomb-Scargle periodogram \citep{VanderPlas_2018}, and the spin period and pulse amplitude are retrieved as observables. Multiple SMC X-ray pulsars regularly occupy the field of view of the RXTE PCA \citep{Yang_2017}. In order to avoid source confusion, the aperiodic X-ray flux is not measured directly by the RXTE PCA. Rather, it is estimated empirically from the pulse amplitude; see Section 2.4.2 and Eq. (5) in \cite{Yang_2017}. \par 
In this paper we analyze the data presented in \cite{Yang_2017}. The time series $P_1(t_n)$ and $L_1(t_n)$ are plotted as grey points in the top two panels of Figure \ref{fig:KFTracking}. Samples whose significance exceeds $99\%$ (defined in Section \ref{sec:SourceProperties}) are overplotted as green points. There are a number of important qualitative and quantitative differences between the synthetic data analyzed by \cite{Melatos_2022} and the real data in Figure \ref{fig:KFTracking} which should be taken into account when comparing the accuracy of the Kalman filter in Section \ref{Sec:KFAnalysis}  with the results in \cite{Melatos_2022}. (i) The fractional fluctuations for SXP 18.3 are approximately three orders of magnitude greater than for the synthetic $P(t)$ data and between one and two orders of magnitude greater than for the synthetic $L(t)$ data; compare the first and second panels in Figure \ref{fig:KFTracking} with Figure 1 in \cite{Melatos_2022}. (ii) Only $8\%$ of the measurements in Figure \ref{fig:KFTracking} (overplotted as green points) are classified by \cite{Yang_2017} as significant pulsations, whereas all $500$ synthetic samples in \cite{Melatos_2022} are implicitly assumed to be significant. (iii)  Between MJD $53750$ and  MJD $55000$, outbursts are visible in the second panel of Figure \ref{fig:KFTracking}, whereas no outbursts are visible in Figure 1 in \cite{Melatos_2022}. Outbursts are common in X-ray timing studies and coincide with many significant detections. The increase in $L_1(t)$ between MJD $53975$ and MJD $54175$ in Figure \ref{fig:KFTracking} is the longest type II outburst recorded for an SMC X-ray pulsar \citep{Schurch_2009}.  
\section{Kalman filter analysis}\label{Sec:KFAnalysis}
A linear Kalman filter \citep{Kalman1960,Gelb_1974} combined with a nested sampler \citep{Speagle_2020} can be applied to the measured time series $P_1(t_n)$ and $L_1(t_n)$ to estimate the posterior distribution of the five fundamental model parameters $\mathbf{\Theta} = (\gamma_\Omega, \gamma_A, \sigma_{AA})$, with $A \in \{ Q,S \}$, conditional on the linear model described by Equations (\ref{eq:EoMTrunc}), (\ref{eq:ME_TruncP}), and (\ref{eq:ME_TruncL}). The Kalman filter likelihood, recursion relations, and nested sampling algorithm are discussed in Section \ref{Sec:KFAlg} and Appendix \ref{sec:PE}.  We discuss the hidden state evolution reconstructed by the Kalman filter in Section \ref{Sec:KFStateTracking}. New estimates of the magnetic dipole moment $\mu$ and radiative efficiency $\eta_0$ of SXP 18.3 are presented in Section \ref{Sec:KFMagMom}. Some consequences of specific systematic uncertainties, e.g.\ due to the misalignment of the magnetic and rotation axes of the star, are discussed in Section \ref{Sec:RadiativeEfficiency} for completeness. The random dispersions of the estimated parameters including $\mu$ and $\eta_0$ are quantified in Section \ref{Sec:Accuracy}.
\subsection{Algorithm}\label{Sec:KFAlg}
In this paper we employ the \texttt{dynesty} nested sampler \citep{Speagle_2020} with the \texttt{bilby} front-end \citep{Ashton_2019} to evaluate the Kalman filter log-likelihood \citep{Meyers_2021}
\begin{equation} \label{eq:LikelihoodMain}
    \ln{p\lr{ \{\bm{Y}_n\}_{n=1}^N |\bm{\Theta}}} = -\frac{1}{2}\sum^N_{n=1}\left[ D_{\bm{Y}} \ln{\lr{2\pi}} + \ln{\textnormal{det}\lr{\bm{s}_n}} + \bm{e}_n^T \bm{s}_n^{-1} \bm{e}_n \right],
\end{equation}
where $\bm{Y}_n = [P_1(t_n), L_1(t_n)]$ is the measurement vector, $D_{\bm Y} = 2$ denotes the dimension of $\bm{Y}_n$, and the innovation vector and its covariance are denoted by $\bm{e}_n$ and $\bm{s}_n = \langle \bm{e}_n, \bm{e}_n^T \rangle$  respectively. In its simplest form, the nested sampling algorithm proceeds as follows. At every step $k$ in the iterative sampling process, the sampler keeps track of $N_{\rm live}$ ``live points'' within the parameter domain, denoted by ${\bf\Theta}^{(k)}_1, \dots, {\bf\Theta}^{(k)}_{N_{\rm live}}$.\footnote{``Dynamic'' nested samplers vary $N_{\rm live}$ throughout the iterative sampling process. At the time of writing, two dynamic nested sampling packages are available with the \texttt{bilby} front-end, namely \texttt{dynesty} \citep{Speagle_2020} and \texttt{PyPolyChord} \citep{Handley_2015}.} The live points are initialized by drawing ${\bf\Theta}^{(1)}_1, \dots, {\bf\Theta}^{(1)}_{N_{\rm live}}$ randomly from the prior distribution $p({\bf\Theta})$. For each $k$, and for all $1 \leq m \leq N_{\rm live}$, we calculate $\mathcal{L}^{(k)}_{m} = \ln{p [\{\bm{Y}_n\}_{n=1}^N|\bm{\Theta}^{(k)}_m]}$, i.e.\ we run the Kalman filter while fixing ${\bf\Theta}^{(k)}_m$ and evaluate Eq. (\ref{eq:LikelihoodMain}). The sampler replaces the live point whose likelihood is lowest, namely $\bm{\Theta}_{m'}^{(k)}$ with $m' = \argmin_{m} \mathcal{L}^{(k)}_{m}$, with a new live point $\bm{\Theta}^{(k+1)}_{m'}$ drawn from the prior $p(\bm{\Theta})$, subject to the condition $\mathcal{L}^{(k+1)}_{m'}> \mathcal{L}^{(k)}_{m'}$. The other live points remain unchanged, viz. $\bm{\Theta}^{(k+1)}_{m} = \bm{\Theta}^{(k)}_{m}$ for $1 \leq m \neq m' \leq N_{\rm live}$. The process repeats until a suitable stopping condition is met, yielding a numerical approximation of the Bayesian evidence and posterior distributions of $\bm{\Theta}$ via weighted histograms or other density estimation techniques. The reader is referred to Sections 5 and 6 in \cite{Skilling_2006} for an overview of the nested sampling algorithm and Table 2 in \cite{Ashton_2022} for a comparison of nested sampling packages.   \par
The Kalman filter generates the pre-fit innovation vector at every time step $t_n$ from
\begin{equation} \label{eq:InnovationVector}
    \bm{e}_n = \bm{Y}_n - \bm{C}\exp[\bm{A}\lr{t_n - t_{n-1}}]\bm{\hat{X}}_{n-1},
\end{equation}
where $\bm{C}$ denotes the $2 \times 3$ observation matrix defined implicitly by the noiseless terms on the right-hand sides of Equations (\ref{eq:ME_TruncP}) and (\ref{eq:ME_TruncL}), and $\bm{A}$ denotes the $3 \times 3$ state transition matrix defined on the right-hand side of Equation (\ref{eq:EoMTrunc}). Replacing $\bm{\hat{X}}_{n-1}$ with the updated state estimate $\bm{\hat{X}}_n$ in Equation (\ref{eq:InnovationVector}) yields the post-fit innovation vector, which is an important metric in estimating the accuracy of the Kalman filter state tracking. The state vector is updated recursively via 
\begin{equation} \label{eq:KFUpdate}
    \bm{\hat{X}}_n = \exp{[\bm{A}\lr{t_n - t_{n-1}}]}\bm{\hat{X}}_{n-1}  + \bm{k}_n \bm{e}_n,
\end{equation}
where $\bm{k}_n$ denotes the Kalman gain. Equations (\ref{eq:InnovationVector}) and (\ref{eq:KFUpdate}) are standard textbook formulas reproduced here for overall context; the reader is referred to Appendix \ref{sec:PE} for further details regarding their structure and implementation.
\subsection{Hidden state tracking}\label{Sec:KFStateTracking}
In Figure \ref{fig:KFTracking} we present the Kalman filter inputs and outputs as functions of time. The time series $P_1(t_n)$ and $L_1(t_n)$ are plotted in the top two panels as grey points and the Kalman filter estimates are overplotted as colored, solid curves. The reconstructed measurements are generated at each time step $t_n$ from the a posteriori state estimates $\hat{\mathbf{X}}_n$ multiplied by $\bm C$, defined in Section \ref{Sec:KFAlg} above. We assess the accuracy with which the Kalman filter tracks the state evolution and hence $P_1(t_n)$ and $L_1(t_n)$ through the average rms error of the post-fit innovation vector using (i) the entire data set, and (ii) the subset of data corresponding to significant detections. In case (i) the average rms errors are $\approx 0.9 \sigma_{P_1}$ and $\approx 0.6 \sigma_{L_1}$, where $\sigma_{P_1}$ and $\sigma_{L_1}$ denote the standard deviations of the entire $P_1(t_n)$ and $L_1(t_n)$  data  set, respectively. In case (ii) the average rms errors are $\approx 1.0\sigma_{P_1}'$ and $\approx 0.8\sigma_{L_1}'$ where $\sigma_{P_1}'$ and $\sigma_{L_1}'$ denote the standard deviations of $P_1(t_n)$ and $L_1(t_n)$ respectively when restricted to the significant detections only. The  results for case (ii) quantify the accuracy with which the Kalman filter tracks the state evolution during outbursts. In the absence of independent measurements of the true hidden state sequence $\bm{X}_n$, the post-fit innovation is a standard metric used in signal processing applications to assess the accuracy of a Kalman filter; see Section 10.1 in \cite{Simon_2006} for an overview on verifying Kalman filter performance. Ultimately, a fuller analysis of accuracy requires $N_{\rm det} > 73$. However, it is encouraging that the Kalman filter partly recovers the two outbursts visible in the second panel of Figure \ref{fig:KFTracking} between MJD 53750 and MJD 55000, even though the idealized state-space model in Section \ref{Sec:LinearDynamics} does not treat outbursts explicitly.  \par 
The Kalman filter state estimates $\bm{\hat{X}}_n = [\hat{\Omega}(t_n),\hat{Q}(t_n),\hat{S}(t_n)]$ are plotted as colored, solid curves in the bottom three panels of Figure \ref{fig:KFTracking}. The shaded regions correspond to the $1\sigma$ state estimate uncertainties returned by the Kalman filter, i.e.\ they are given by the square root of the diagonal elements in Equation (\ref{Eq:KFCovariance}).  The mean-reverting nature of the hidden state variables is clear upon inspection. The components $\Omega_0, Q_0,$ and $S_0$ of the state vector in equilibrium  are overplotted as black, dashed lines. We calculate the sample mean $\Omega_0$ from Equation (A1) in \cite{Melatos_2022} and infer $Q_0$ and $S_0$ from the mode of the posterior of $\gamma_\Omega$, discussed in Section \ref{sec:MagMomentObs}. The accretion rate fluctuations increase from $  | Q_1(t) | Q_0 \lesssim 10^{-9} M_\odot \; \textnormal{yr}^{-1}$ during quiescence to $| Q_1(t) | Q_0 \lesssim 10^{-8} M_\odot \; \textnormal{yr}^{-1}$ during outburst which is broadly consistent with other sources; see Figure 3 in \cite{Mukherjee_2018} for example. The Maxwell stress fluctuations satisfy $ | S_1(t) | S_0 \lesssim  10^7 \, \textnormal{g}\, \textnormal{cm}^{-1} \, \textnormal{s}^{-2}$. They correlate somewhat with the pulse period and the aperiodic X-ray flux in general and during outbursts in particular; we measure the Pearson $r$-coefficients as $r[S_1(t),L_1(t)] \approx 0.19 \pm 0.034$ and $r[S_1(t),P_1(t)] \approx -0.36 \pm 0.032$, where the uncertainties correspond to the standard error on $r$. We also find evidence of a moderate correlation between the Kalman filter output for $S_1(t_n)$ and $Q_1(t_n)$, with $r[S_1(t),Q_1(t)] \approx 0.26 \pm 0.033$. An instance of the correlation is visible in Figure \ref{fig:KFTracking} between MJD 53975 and MJD 54175; the magnetosphere is compressed relative to equilibrium, with $Q(t) > Q_0$, causing spikes in $S(t)$. The results about $S_1(t)$ are examples of the new and physically important information that flows from a Kalman filter analysis. \par 
In the bottom panel of Figure  \ref{fig:KFTracking},  $S\lr{t}$ goes negative near MJD 51700, MJD 53200, MJD 53400, and MJD 54600. Negative excursions of $S(t)$ are unphysical, but they are permitted in principle (albeit rarely) by the Ornstein-Uhlenbeck dynamics in (\ref{eq:LE_Q}) and (\ref{eq:LE_S}) in the linear regime; see Section 2.4 in \cite{Melatos_2022}, where the possibility is discussed. The negative excursions of $S(t)$ are brief, i.e.\ 0.8 \% of the total observation time. Enforcing $S(t)>0$ artificially for all $t$ does not change the state estimates appreciably.\footnote{In a more general accretion model, where $S(t)$ is proportional to the product of the toroidal and poloidal magnetic field components, both signs of $S(t)$ are possible \citep{Wang_1987,Dangelo_2010,Dangelo_2012}.} 
\subsection{Magnetic moment and radiative efficiency}\label{Sec:KFMagMom}
Figure \ref{fig:CornerPlot} presents the five-dimensional posterior distribution of the model parameters $\mathbf{\Theta} = \lr{\gamma_\Omega, \gamma_A, \sigma_{AA}}$, with $A \in \left\{ Q, S \right\}$, returned by the nested sampler and visualised in cross section through a traditional corner plot. The parameter estimates are reported at the top of each one-dimensional posterior (histogram; $\log_{10}$ scale). We quantify the uncertainty in the estimate using a $68\%$ credible interval, visible as three, vertical, dashed lines in the one-dimensional posteriors, and through the innermost contour in the two-dimensional posteriors in Figure \ref{fig:CornerPlot}. The nominal value corresponds to the posterior median. The mode (peak) of each posterior is contained within the $68\%$ credible interval, and is employed as a point estimate to generate the state sequence $\bm{\hat{X}}_n$ plotted in Figure \ref{fig:KFTracking}. The steps required to reproduce the results in Figures \ref{fig:KFTracking} and \ref{fig:CornerPlot} are outlined in Section 3.1 in \cite{Melatos_2022}. We assume $\eta\lr{t} = \eta_0$ in the present paper, thereby neglecting the parameters $\gamma_\eta$ and $\sigma_{\eta\eta}$ in the Ornstein-Uhlenbeck equation for $\eta(t)$; see Equation (11) in \cite{Melatos_2022}.  \par
In Figure \ref{fig:CornerPlot}, four parameters are estimated unambiguously, namely $\gamma_\Omega, \gamma_Q, \gamma_S$, and $\sigma_{QQ}$. There is evidence of $\gamma_\Omega$-$\gamma_S$ and $\gamma_Q$-$\sigma_{QQ}$ correlations, visible as a diagonal tilt in the contours in the $\gamma_\Omega$-$\gamma_S$ and $\gamma_Q$-$\sigma_{QQ}$ planes. In contrast, we observe railing in the posterior for $\sigma_{SS}$, which cannot be estimated unambiguously. Although the  identifiability analysis presented in Appendix D in \cite{Melatos_2022} suggests that all five model parameters including $\sigma_{SS}$ should be identifiable with sufficient data, the estimation problem in this paper using real observational data is more challenging than the synthetic exercise in \cite{Melatos_2022}, e.g.\ compare the second panel in Figure 1 in \cite{Melatos_2022} with Figure \ref{fig:KFTracking} below. Figure \ref{fig:CornerPlot} suggests that one may require $N > 854$ for SXP 18.3 to constrain all five model parameters. \par  
Figure \ref{fig:CornerPlotInferred} presents the posterior of the four inferred parameters $(Q_0, S_0, \eta_0, \mu)$ and is visualised in cross section through a traditional corner plot. To generate the histograms
we draw random samples from the marginalized posterior of $\gamma_\Omega$ and the normal distribution $A \sim \mathcal{N}(A_0, \sigma_A N^{-1/2})$, with $A \in \{\Omega, L\}$, where $\sigma_A N^{-1/2}$ denotes the standard error of the sample means, $\Omega_0$ and $L_0$.  The strong pairwise covariances (diagonal contours) observed in Figure \ref{fig:CornerPlotInferred} follow directly from the ratios in Equations (A4) and (A6) in \cite{Melatos_2022} and Equation (\ref{Eq:S0_denominator}) above. The parameter estimates inferred in this paper are reported in the first row in Table \ref{tab:EqParams}. We find $Q_0 = 2.4^{+0.8}_{-0.7} \, \times \, 10^{17} \, {\, \rm \, g \, s^{-1}}, \, S_0 = 5.6^{+2.0}_{-1.6} \, \times \, 10^6 \, {\rm g \, cm^{-1} \, s^{-2}}, \, \eta_0 = 0.04^{+0.02}_{-0.01}$,  and hence $\mu = 8.0^{+1.3}_{-1.2} \, \times 10^{30} \, \, {\rm G \, cm^3}$, compared to $\mu = 5.0^{+1.0}_{-1.0}\, \times \, 10^{30} \, {\rm G \, cm^3}$ as inferred traditionally from time-averaged data assuming $Q_0 \propto \langle L(t) \rangle$ and $\eta_0 = 1$ (bottom row of Table \ref{tab:EqParams}).  The latter, traditional approach implies that $100\%$ of the gravitational potential energy of material falling onto the stellar surface is converted to heat and hence X-rays. It is a standard assumption in X-ray timing analysis. The results presented in this paper suggest otherwise, with $\eta_0 \neq 1$ in general, i.e.\ we find that only  $\approx 4\%$ of the specific gravitational potential energy of infalling matter in SXP 18.3 is converted to X-rays. As a result, the Kalman filter estimates of $Q_0$ and hence $\mu$ are higher than the estimates based on time-averaged data and $\eta_0 = 1$. The present analysis yields independent measurements of $\eta_0$ and hence $\mu$ in the X-ray pulsar context and is another example of a new and astrophysically important result which stems from the Kalman filter framework. 
\par 
We draw the reader's attention to the following point of interest. In \cite{Klus_2014} and \cite{Dangelo_2017}, the authors estimated $B$ and hence $\mu$ for 38 SMC X-ray pulsars. Assuming magnetocentrifugal equilibrium, \cite{Klus_2014} found that $\approx 50\%$ of the stars have $B \geq 4.4\times 10^{13} \, {\rm G}$, i.e.\ above the Schwinger limit and comparable to magnetar field strengths, based on $Q_0 \propto \langle L(t) \rangle$ and $\eta_0 = 1$. However, one must be cautious about interpreting $B$ at face value. \cite{Dangelo_2017} argued that the $Q_0$ estimates in \cite{Klus_2014} correspond to the average mass accretion rate during outburst and suggested  $Q_0^* = \lr{N_{\textnormal{det}}/N} Q_0$ as a proxy for the mass accretion rate during quiescence, where the ratio $N_{\textnormal{det}}/N$ approximates the fraction of time spent in outburst. Adopting $Q_0^*$ instead of $Q_0$ yields $\approx 15\%$ of SMC X-ray sources with magnetar field strengths; see Figure 22 in \cite{Dangelo_2017} for a comparison of $B$ estimates using $Q_0$ and $Q_0^*$, plotted as green triangles and cyan points respectively. Similarly, the free parameter $\gamma_\Omega$ and hence the components $Q_0,S_0,$ and $\eta_0$ estimated in this paper are influenced by episodes of quiescence and type II outbursts \citep{Reig_2008,Franchini_2021}; see the second panel in Figure \ref{fig:KFTracking}. Following \cite{Dangelo_2017}, we report two values for the magnetic dipole moment in the first and second rows in Table \ref{tab:EqParams}. The second value, $\mu^\ast = 2.3^{+0.4}_{-0.3} \times 10^{30} \, {\rm G \, cm^3}$, corresponds to the inferred magnetic dipole moment including the $N_{\rm det}/N$ correction for outbursts \citep{Dangelo_2017}. 
\subsection{Impact of model assumptions}\label{Sec:RadiativeEfficiency}
Systematic model uncertainties affect the inferred values of $\mu$ and $\eta_0$ in complicated ways, which are hard to quantify with the data at hand. \cite{Yatabe_2018} discussed systematic uncertainties thoroughly in Section 4 of the latter reference, with an emphasis on the canonical picture of magnetocentrifugal accretion. For example, disk warping and precession induced by misalignment of the magnetic and rotation axes \citep{Foucart_2011,Romanova_2021} cause fluctuations in the mass accretion rate $Q(t)$ \citep{Romanova_2021}. If $Q(t)$ and $\eta(t)$ are anticorrelated, spikes in $Q(t)$ produce dips in $\eta(t)$, affecting the inferred value of $\eta_0$. Such effects have been investigated recently in Figures 6 and 11 in \cite{Romanova_2021}. Misalignment of the rotation and magnetic axes leads to complicated three-dimensional accretion flows \citep{Romanova_2021,Melatos_2022}, requiring sophisticated, three-dimensional magnetohydrodynamic simulations, which lie outside the scope of the current analysis. Other $\eta_0$ and $\mu$ biases neglected in this paper arise due to uncertainties in the location of the magnetospheric radius $R_{\rm{m}}$, and X-ray beaming, among other factors.  The accretion dynamics in Section \ref{Sec:Torque} are presented as a compromise between realism and the explanatory power of the limited volume of data that is currently available, i.e.\ $\sim 10^3$ samples. Disentangling these and other complicated uncertainties will require larger datasets from future X-ray timing experiments. We emphasize that the Kalman filter approach complements (and does not supplant) other approaches to measuring $\mu$ \citep{Klus_2014,Ho_2014,shi_2015,Christodoulou_2016}, which also neglect some of the above systematics but are likely to depend on them in different ways. Applying multiple approaches helps to assemble a complete picture. \par
In the current single-object analysis, it is difficult to ascertain what properties of SXP 18.3 dictate the inferred value $\eta_0=0.04^{+0.02}_{-0.01}$. For example, one may ask: does the information that determines $\eta_0$ come mainly from $P_1(t)$ or $L_1(t)$ fluctuations? In rough terms, the answer is both: $P_1(t)$ implies higher $Q_1(t)$ than would be inferred assuming $\eta_0=1$ traditionally, implying $\eta_0 \ll 1$ to match $L_1(t)$ as observed. \par 
To get a preliminary sense of how $P_1(t)$ and $L_1(t)$ interact to produce the inferred $\eta_0$ value, we boost $L_{1}(t)$ artificially to test how $\eta_0$ changes. The results of the test are presented in Figure \ref{fig:InflateMeasurement}, where we plot $\eta_0$ (vertical axis), inferred using a different set of $L_1(t)$ fluctuations artificially enhanced by a factor $\tilde{L}$, versus $\tilde{L}$ (horizontal axis). The numerical experiment follows the same procedure outlined in Section \ref{Sec:KFAlg}. It reveals that $\eta_0$ increases, as $\tilde{L}$ increases, i.e.\ higher $L_1(t)$ fluctuations yield higher $\eta_0$, with $\eta_0 \propto \tilde{L}^{1/2}$ for $\tilde{L} \gtrsim 1$. The opposite applies to $P_1(t)$: $\eta_0$ decreases, as $\tilde{P}$ increases (not plotted for brevity).  A more elaborate physical understanding in terms of the ``moving parts'' of magnetocentrifugal accretion is likely to emerge upon extending the present analysis to a larger population of X-ray pulsars. In particular, instantaneous (anti)correlations between the state variables and measurements, such as those presented in Section \ref{Sec:KFStateTracking} for SXP 18.3, are likely to shed light on physical causes and their effects at the disk-magnetosphere boundary once measured at the population level. We defer this substantial project to a forthcoming paper.
\par
One possible astrophysical interpretation of $\eta_0=0.04^{+0.02}_{-0.01}$ (out of many) is that disk accretion, as modelled in Section \ref{Sec:Torque}, is supplemented by outflows, as revealed by three-dimensional magnetohydrodynamic simulations \citep{Romanova_2015}, e.g.\ slingshot-driven by the corotating magnetosphere in the regime $R_{\rm{c}} \lesssim R_{\rm{m}}$. Some qualitative features of outflows are captured fortuitously by the mean-reverting disk dynamics in Equations (\ref{eq:LE_Q}) and (\ref{eq:LE_S}); see the fourth and fifth panels of Figure \ref{fig:KFTracking} for time-resolved histories of $Q(t)$ and $S(t)$, revealing episodes of quiescence punctuated by accretion outbursts, as well as Appendix C.4 in \cite{Melatos_2022} for an abridged review of outflow-launching mechanisms. It is plausible, in this context, that outflows are stronger (and hence $\eta_0$ is lower) when $Q_0$ is higher, as pressure builds up near the disk-magnetosphere boundary. It will be interesting to test for the existence of an inverse relation between $\eta_0$ and $Q_0$ in a future Kalman filter analysis of the SMC HMXB population.
\subsection{Accuracy}\label{Sec:Accuracy}
Without independent measurements of the true hidden state sequence ${\bf X}_n = [\Omega(t_n), Q(t_n), S(t_n), \eta(t_n)]$, it is challenging to verify how accurately the Kalman filter state estimates $\hat{\bf X}_n$ reproduce ${\bf X}_n$ in an astrophysical setting, e.g.\ SXP 18.3. \cite{Melatos_2022} conducted Monte Carlo validation tests on the linear problem defined by Equations (13)--(16) in \cite{Melatos_2022}, using synthetic time series with $N=500$, i.e.\ comparable in volume to the SXP 18.3 time series. The tests return an error in the peak of the $\gamma_\Omega$ posterior of $\approx 0.15 \, {\rm dex}$ relative to the injected value, comparable to the uncertainty in the posterior for $\gamma_\Omega$ in Figure \ref{fig:CornerPlot}. \par 
The dispersion introduced by randomness in the noise realizations also limits the accuracy of the parameter estimates, because one cannot know where the actual, astrophysical noise realization lies within the ensemble of possibilities. \cite{Melatos_2022} found that the full width half maximum (FWHM) of the $\gamma_\Omega$ posterior in validation tests with synthetic data equals $\approx 0.60 \, {\rm dex}$, which agrees approximately with the FWHM in Figure \ref{fig:CornerPlot} ($\approx 0.31 \,{\rm dex}$). Monte Carlo validation tests also yield FWHMs of the posteriors of $\gamma_Q$, $\gamma_S$, $\sigma_{QQ}$, and $\sigma_{SS}$ in line with the results in Figure \ref{fig:CornerPlot}, e.g.\ the FWHM of the $\gamma_Q$ posterior is $\approx 0.60 \, {\rm dex}$ and $\approx 0.24 \, {\rm dex}$ in the validation tests in \cite{Melatos_2022} and Figure \ref{fig:CornerPlot}, respectively. \par
\begin{figure}
\centering{
    \includegraphics[width=0.95\textwidth, keepaspectratio]{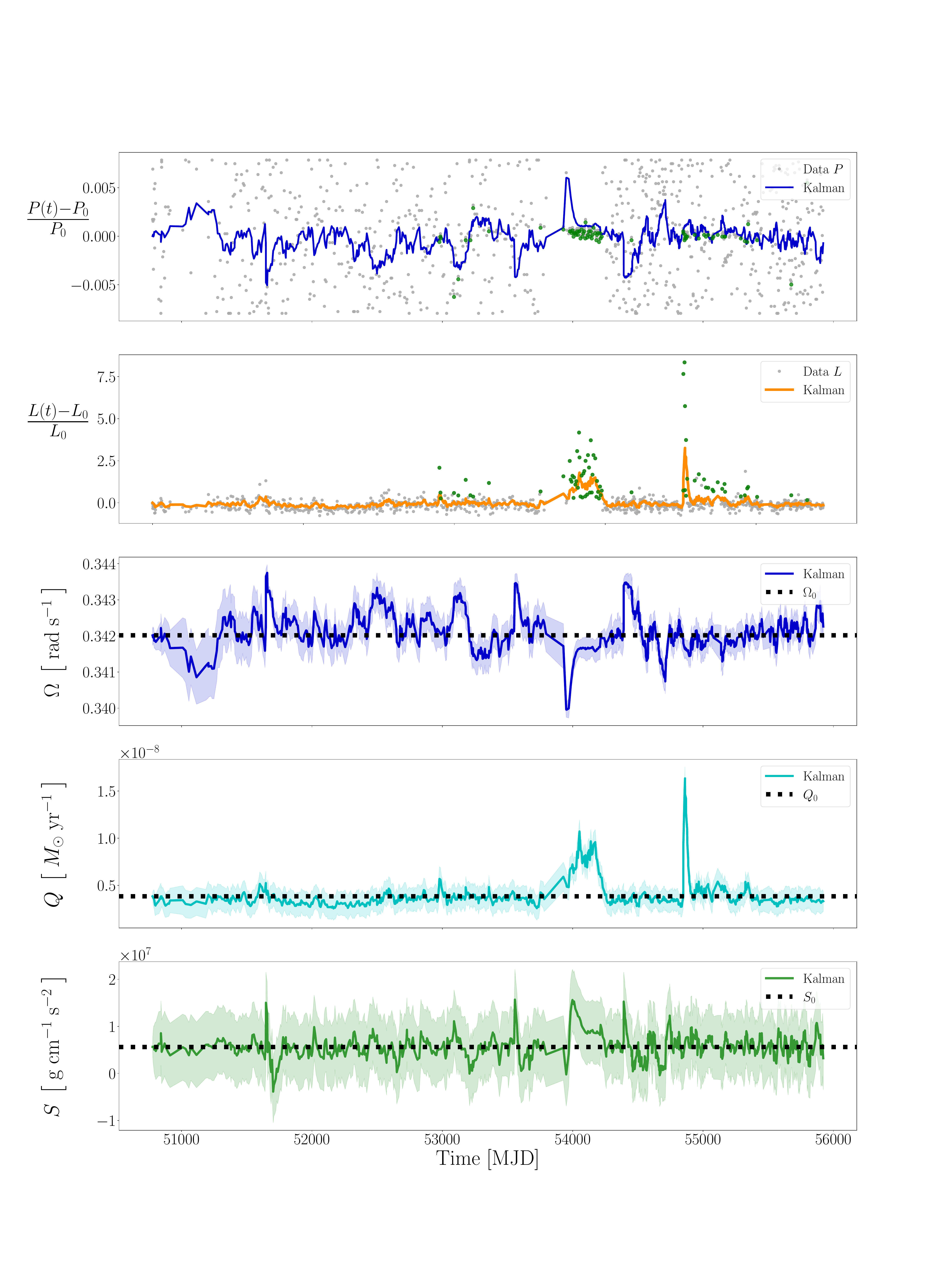}}
        \caption{Kalman filter state tracking applied to SXP 18.3. Inputs: observational data of fractional pulse period fluctuations $P_1\lr{t}$ (first panel) and aperiodic X-ray luminosity fluctuations $L_1\lr{t}$ (second panel). The RXTE data are plotted as grey points, with significant pulsations overplotted as green points (first and second panels). Outputs: Kalman filter estimates $\hat{P}_1\lr{t}$ (top panel, blue curve) and $\hat{L}_1\lr{t}$ (second panel, orange curve), and the absolute (not fractional) state variables $\Omega\lr{t}$ (third panel, blue curve), $Q\lr{t}$ (fourth panel, cyan curve) and $S\lr{t}$ (fifth panel, green curve). In the bottom three panels, the state vector equilibrium parameters are plotted as horizontal, dashed lines. The shaded regions correspond to the $1\sigma$ state estimate uncertainties, returned by the Kalman filter. $\Omega_0$ and $Q_0$ are inferred from Equations (A1) and (A4) in \cite{Melatos_2022} and $S_0$ is inferred from Equation (\ref{Eq:S0_denominator}) above, discussed in Section \ref{sec:MagMomentObs}. The time units on the horizontal axis are MJD. }
    \label{fig:KFTracking}
\end{figure}

\begin{figure}
\centering{
\hspace*{-2.5cm}
\includegraphics[width=1.3\textwidth, keepaspectratio]{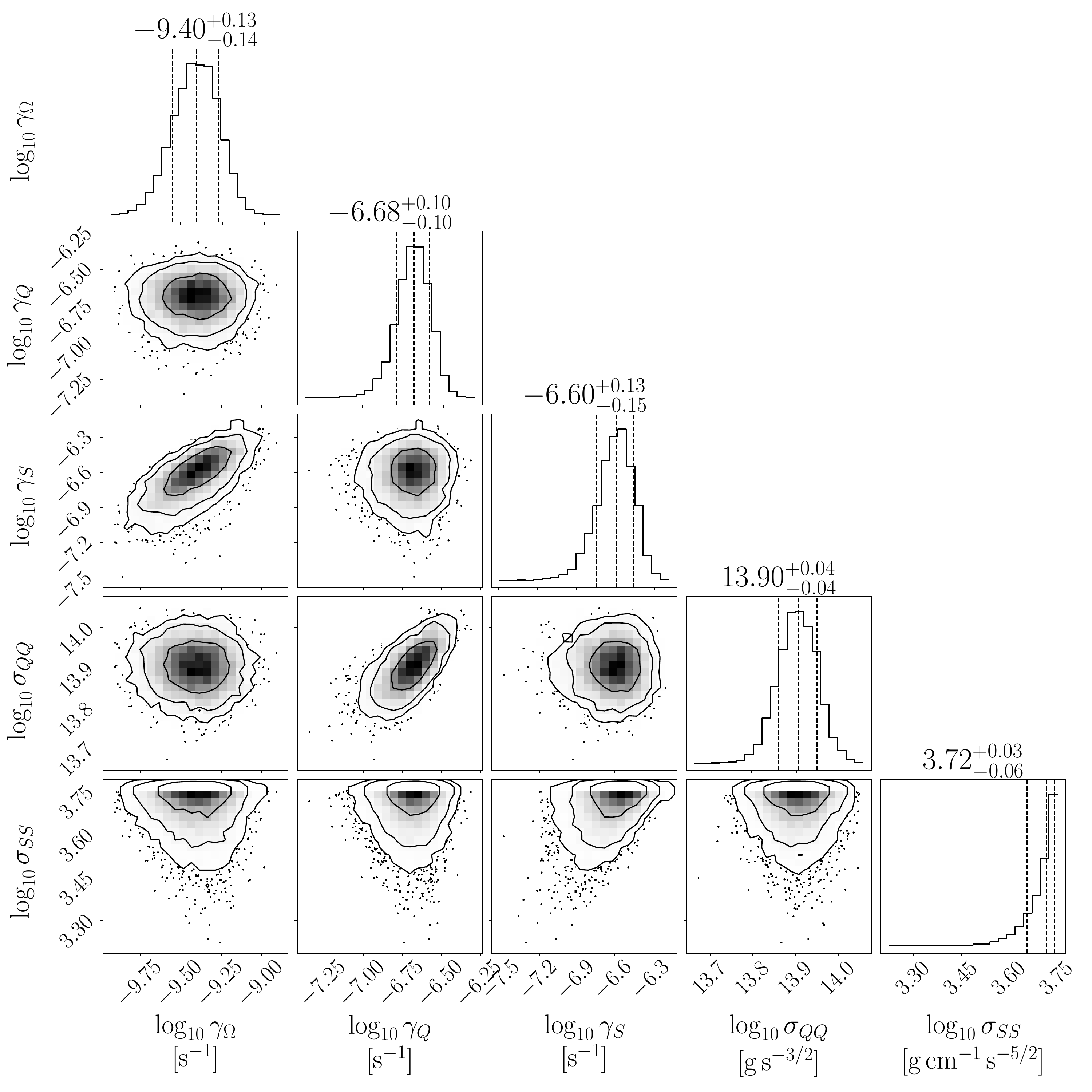}}
    \caption{Corner plot of the posterior distribution of the five fundamental model parameters $\mathbf{\Theta}$ = $(\gamma_\Omega, \gamma_A, \sigma_{AA})$, with $A \in \left\{ Q, S \right\}$, for SXP 18.3. All distributions are plotted on a $\log_{10}$ scale. Contour plots depict the posterior marginalized over three out of five parameters. Histograms depict the posterior marginalized over four out of five parameters. We estimate four parameters unambiguously, namely $\gamma_\Omega, \gamma_Q, \gamma_S,$ and $\sigma_{QQ}$. In contrast, $\sigma_{SS}$ rails against the upper bound of the prior and cannot be recovered. }
    \label{fig:CornerPlot}
\end{figure}

\begin{figure}
\centering{
\hspace*{-2.5cm}
\includegraphics[width=1.3\textwidth, keepaspectratio]{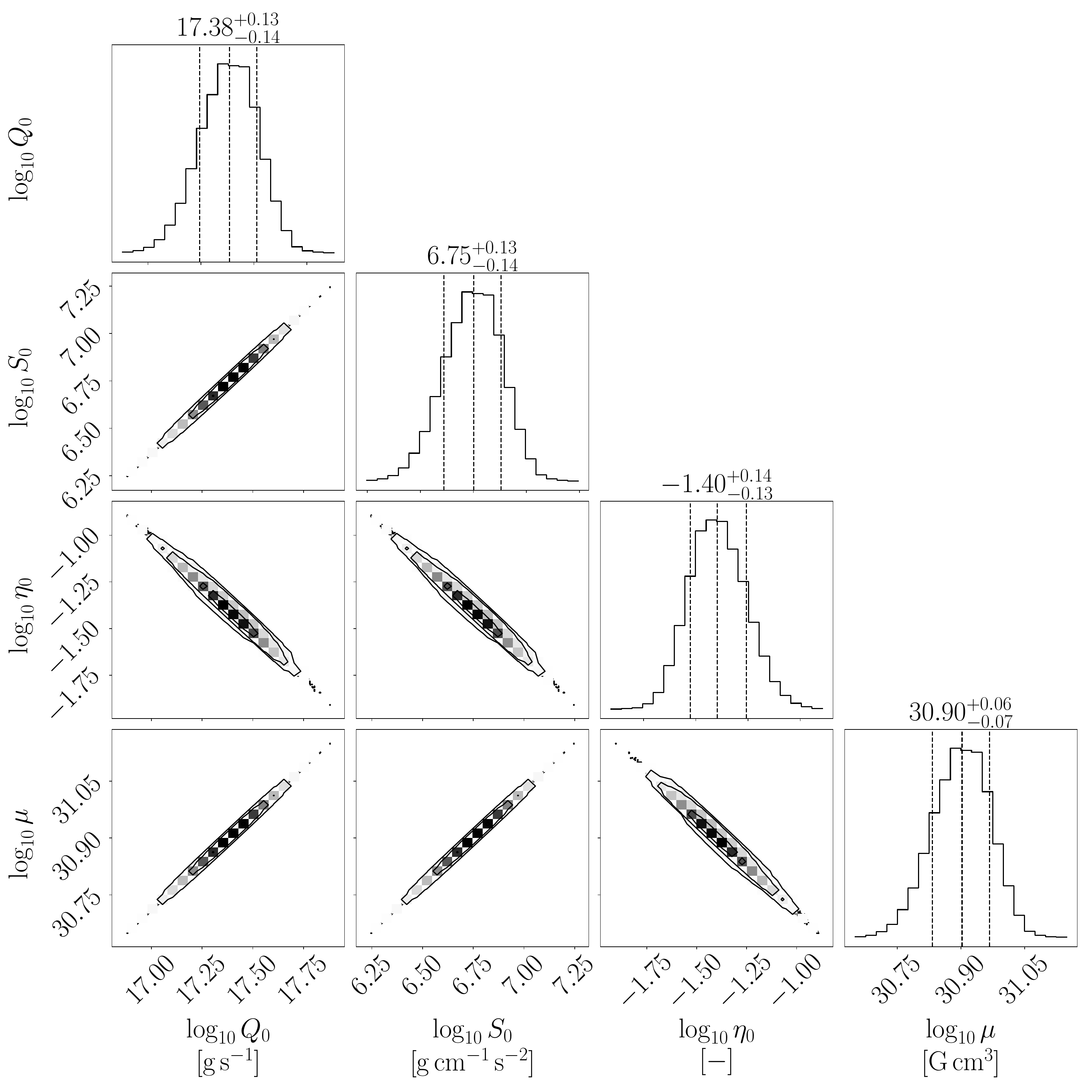}}
    \caption{Corner plot of the posterior distribution of the four induced model parameters $(Q_0,S_0,\eta_0,\mu)$ for SXP 18.3. All distributions are plotted on a $\log_{10}$ scale. Contour plots depict the posterior marginalized over two out of four parameters. Histograms depict the posterior marginalized over three out of four parameters.}
    \label{fig:CornerPlotInferred}
\end{figure}

\begin{figure}
\centering{
\hspace*{-1.5cm}
    \includegraphics[width=0.9\textwidth, keepaspectratio]{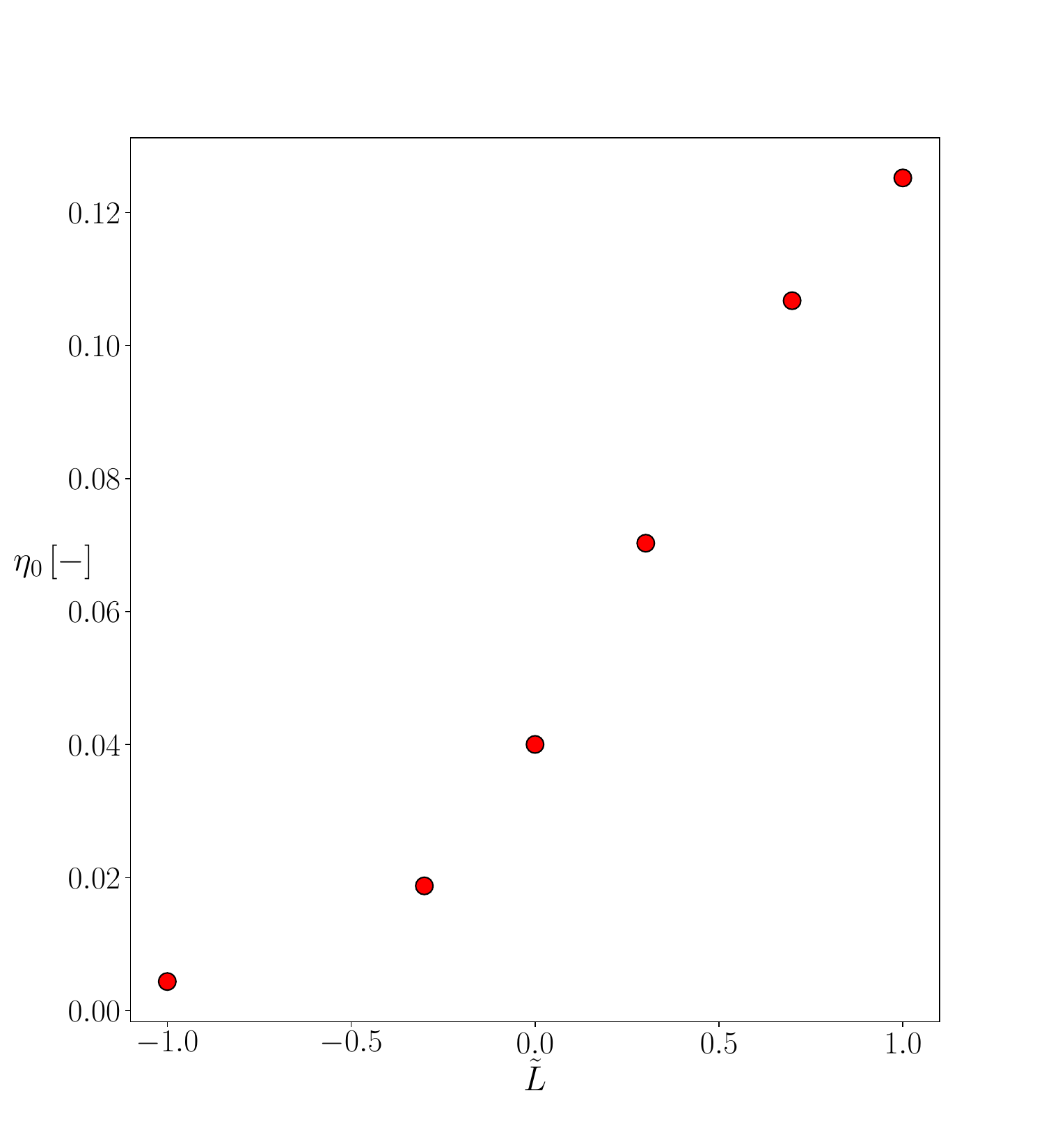}}
    \caption{Radiative efficiency $\eta_0$ inferred using a different set of $L_1(t)$ fluctuations artificially enhanced by a factor $\tilde{L}$, versus $\tilde{L}$. The plotted points correspond to the $\eta_0$ one-dimensional posterior median values. The horizontal axis is plotted on a $\log_{10}$ scale.}
    \label{fig:InflateMeasurement}
\end{figure}

 
The analysis yields new estimates of the mean-reversion time-scales $\gamma_Q$ and  $\gamma_S$, as well the rms noise amplitude of the accretion rate. We find $\gamma_Q = 2.1^{+0.5}_{-0.4} \, \times 10^{-7} \, {\rm s^{-1}}$, $\gamma_S = 2.5^{+0.9}_{-0.7} \, \times 10^{-7} \, {\rm s^{-1}}$, and $\sigma_{QQ} = 8.0^{+0.8}_{-0.8} \times 10^{13} \, {\rm g \, s^{-3/2}}$. Independent estimates of $\gamma_A$ and $\sigma_{AA}$ with $A \in \{ Q, S \}$ are not available at the population level, e.g.\ for all SMC X-ray pulsars, because traditional analysis methods based on time-averaged data do not estimate $\gamma_A$ and $\sigma_{AA}$, as discussed in Section \ref{sec:intro}. However, the results in Figure \ref{fig:CornerPlot} should be regarded as reasonable for two qualitative reasons. First, the characteristic timescale of mean reversion in the bottom two panels in Figure \ref{fig:KFTracking} and the second and third columns in Figure \ref{fig:CornerPlot} is $\gamma_{A}^{-1} \sim 10^{7} \rm{\;s}$, with $A \in \{Q,S\}$. This is consistent with the Lomb-Scargle PSD computed from the light curves of other accreting systems such as Scorpius X$-$1 and 2S 1417$-$624, which roll over at $\sim 10^{-7} \rm{\; s}^{-1}$ \citep{Mukherjee_2018,Serim_2022}. Second, the inferred accretion rate fractional fluctuations satisfy $\sigma_{QQ} \approx 0.7Q_0 \gamma_Q^{1/2}$ on average, in line with $\sigma_{QQ} \approx 0.4Q_0 \gamma_Q^{1/2}$ in Figure 3 in \cite{Mukherjee_2018}.
\begin{table}
\centering
\setlength{\tabcolsep}{5pt}
\hspace*{-2.75cm}
 \begin{tabular}{c c c c c} 
 \hline
  Analysis & $Q_0 \, [10^{17} \, {\rm g \, \rm s^{-1}}]$  & $S_0 \, [10^6 \, {\rm g \, cm^{-1} \, s^{-2}}]$ & $\eta_0 \, [-]$ & $\mu \, [10^{30} \, {\rm G \, cm^3}]$ \\ [1ex] \hline 
  Kalman filter & $2.4^{+0.8}_{-0.7}$ & $5.6^{+2.0}_{-1.6}$ &  $0.04^{+0.02}_{-0.01}$ &  $8.0^{+1.3}_{-1.2}$ 
 \\ [1ex] Outburst correction &
    $0.2^{+0.07}_{-0.05}$ & $-$  & $-$ & $2.3^{+0.4}_{-0.3}$ \\  [1ex]
\cite{Klus_2014} & $-$ & $-$  & $1.0$  & $5.0^{+1.0}_{-1.0}$ \\  [1ex]
 \hline
 \end{tabular}
 \caption{Equilibrium quantities $Q_0$ (second column), $S_0$ (third column), and $\eta_0$ (fourth column), together with the magnetic dipole moment $\mu$ (fifth column). The first row corresponds to the linear Kalman filter analysis in Sections \ref{sec:AD} and \ref{Sec:KFAnalysis} above. The second row corresponds to $Q_0^\ast$ and $\mu^\ast$, with a correction made for outbursts \citep{Dangelo_2017}. The third row corresponds to using time-averaged data assuming $Q_0 \propto \langle L(t) \rangle$ and $\eta_0 = 1$ \citep{Klus_2014}.}
 \label{tab:EqParams}
\end{table}
%
\section{Conclusion}\label{Sec:Conclusions}
Observations of accretion-powered pulsars in the SMC \citep{Galache_2008,Coe_2015,Yang_2017} and elsewhere reveal stochastic variability in the measured aperiodic X-ray flux $L(t)$ and pulse period $P(t)$ \citep{Bildsten_1997}, driven by several complicated mechanisms such as hydromagnetic instabilities at the disk-magnetosphere boundary \citep{Romanova_2003,Kulkarni_2008,Das_2022} and flicker noise due to propagating fluctuations in the disk $\alpha$ parameter \citep{Lyubarskii_1997}. The time series $L(t_n)$ and $P(t_n)$ can be combined in several ways to shed light on the accretion physics at the disk-magnetosphere boundary and important stellar properties, e.g. the magnetic dipole moment $\mu$. Most previous, pioneering work in this direction involves employing the time-averaged observables $L_0 = \langle L(t) \rangle$ and $ P_0 = \langle P(t) \rangle$ coupled with predictions from physical theories of accretion \citep{Klus_2014, Ho_2014,Mukherjee_2015,shi_2015, Dangelo_2017,karaferias_2023} and a priori assumptions about the rotational state of the star, e.g. magnetocentrifugal equilibrium, to infer $\mu$. The aforementioned references share a common feature, i.e.\ they assume a value for $\eta_0$, typically $\eta_0 = 1$, which cannot be inferred uniquely from time-averaged data only. Accordingly, one must be careful about interpreting $\mu$, when $\eta_0$ is prescribed arbitrarily. Moreover, the time series $P(t_n)$ and $L(t_n)$ contain an abundance of important time-dependent information such as instantaneous (anti)correlations between state variables and measurements, which is lost after averaging over time.   \par
In this paper, we  employ the Kalman filter framework developed by \cite{Melatos_2022} and apply it to {\em RXTE} data for the SMC X-ray transient SXP 18.3, breaking the degeneracy between $\mu$ and the radiative efficiency. The analysis sheds light on the efficiency of radiative processes near the disk-magnetosphere boundary for accretion-powered pulsars, returning $\eta_0 = 0.04^{+0.02}_{-0.01}$, in contrast with the standard assumption $\eta_0 = 1$. The analysis also returns $\mu = 8.0^{+1.3}_{-1.2} \times 10^{30} \, {\rm G \, cm^{3}}$, to be compared with $5.0^{+1.0}_{-1.0} \times 10^{30} \, {\rm G \, cm^{3}}$ assuming $\eta_0 = 1$ in \cite{Klus_2014}. That is, $\eta_0$ and $\mu$ are estimated independently by exploiting temporal information in the fluctuations $L_1(t_n)$ and $P_1(t_n)$. \par  
The Kalman filter reveals new information about how $S(t)$, the Maxwell stress at the disk-magnetosphere boundary, fluctuates with time. The hidden variable $S(t)$ is interesting astrophysically and notoriously difficult to measure. We find moderate evidence of (anti)correlations between the Kalman filter output for $S_1(t)$ and the observables $P_1(t)$ and $L_1(t)$, with $r[S_1(t),L_1(t)] \approx 0.19 \pm 0.034$ and $r[S_1(t),P_1(t)] \approx -0.36 \pm 0.032$. Similarly, we find evidence of correlations between the Kalman filter output for $S_1(t)$ and $Q_1(t)$, with $r[S_1(t),Q_1(t)] \approx 0.26 \pm 0.033$. The latter correlation is visible in Figure \ref{fig:KFTracking} between MJD $53975$ and MJD $54175$; when the magnetosphere is compressed relative to equilibrium, i.e.\ $Q(t) > Q_0$, we observe spikes in $S(t)$. \par  
The analysis yields new estimates for the mean-reversion time-scales $\gamma_Q = 2.1^{+0.5}_{-0.4} \, \times 10^{-7} \, {\rm s^{-1}}$ and $\gamma_S = 2.5^{+0.9}_{-0.7} \, \times 10^{-7} \, {\rm s^{-1}}$, together with the rms noise amplitude of the accretion rate, $\sigma_{QQ} = 8.0^{+0.8}_{-0.8} \times 10^{13} \, {\rm g \, s^{-3/2}}$. The inferred fluctuation parameters, $\gamma_Q, \gamma_S,$ and $\sigma_{QQ}$ are consistent with other X-ray pulsars, e.g. the PSD of $P(t_n)$ fluctuations measured for 2S 1417$-$624 rolls over at $\sim 10^{-7} \, {\rm s^{-1}}$; see Figure 6 in \cite{Serim_2022}.    \par 
The next step is to apply the Kalman filter parameter estimation scheme to the entire catalogue of accretion-powered pulsars analyzed by \cite{Yang_2017}. The catalogue consists of $30$ stars near magnetocentrifugal equilibrium, with $\tn{d}P/\tn{d}t \approx 0$, and 23 stars existing in a state of magnetocentrifugal disequilibrium, with $\tn{d}P/\tn{d}t \not\approx 0$, i.e. spinning up or down secularly. The former subsample can be treated in principle using the linear Kalman filter and state and measurement models presented in \cite{Melatos_2022} and discussed in Sections \ref{sec:AD} and \ref{Sec:KFAnalysis} above. The latter subsample, however, poses a new challenge and necessitates three adjustments to the current framework. First, we replace Equations (\ref{eq:ME_TruncP}) and (\ref{eq:ME_TruncL}) with their nonlinear counterparts, Equations (\ref{eq:SpinPeriod1}) and (\ref{eq:Luminosity1}), respectively. Second, we replace Equation (\ref{eq:EoMTrunc}) with the nonlinear magnetocentrifugal torque law  (\ref{eq:spinequation1}), and a system of modified Langevin equations; see Equations (B1)--(B3) in \cite{Melatos_2022}. Third, we replace the linear Kalman filter with a nonlinear filter, e.g. an unscented or extended Kalman filter. Estimating $\mu$, $\eta_0$, and instantaneous (anti)correlations between time series at a population level may shed light on accretion physics in the SMC in general. We will explore this opportunity in a forthcoming paper. 
\par
The authors thank Julian Carlin, Liam Dunn, Tom Kimpson, and Andr\'es Vargas for helpful discussions about X-ray timing data and pulsar astrophysics. We also acknowledge a helpful discussion with Robin Evans regarding Kalman filter accuracy. We thank the anonymous referee for constructive feedback which improved the manuscript. This research was supported by the Australian Research Council Centre of Excellence for Gravitational Wave Discovery (OzGrav), grant number CE170100004. NJO’N is the recipient of a Melbourne
Research Scholarship. DMC, SB, and STGL acknowledge funding through the National Aeronautics and Space Administration Astrophysics Data Analysis Program grant NNX14-AF77G.
\bibliographystyle{aasjournal}
\bibliography{main.bib}
\appendix
\section{Kalman Filter Parameter Estimation} \label{sec:PE}
In this appendix, we summarise the Kalman filter algorithm in general for  an arbitrary linear dynamical system given a set of noisy observations. The Kalman filter recursion relations and associated likelihood are written down in Appendix \ref{App:KFState} and \ref{App:ParameterEstimation} respectively. Although the material in this appendix appears in many classic textbooks \citep{Gelb_1974,Shumway_2000,Challa_2011}, it is summarized here briefly for the convenience of the reader in a notation that is compatible with the body of the paper, to help the reader reproduce the results in Section \ref{Sec:KFAnalysis}. 
\subsection{State tracking}\label{App:KFState}
The Kalman filter is a powerful signal processing tool, providing estimates of non-observable, hidden state variables $\bm{X}$ given a set of noisy measurements $\bm{Y}$. It is a sequential estimation algorithm, in the sense that the epoch of the estimated state is updated with each new observation, and all information obtained from previous measurements is contained in the current state and covariance estimates denoted by $\bm{\hat{X}}$ and $\bm{P}$ respectively. \par
In this paper, we consider a system of first order, noise-driven, linear differential equations of the following form 
\begin{equation} 
        \frac{\textnormal{d} \bm{X}}{\textnormal{d} t} = \bm{A}\bm{X}\left(t\right) + \bm{\xi}\left(t\right), \label{eq:generalODE}
\end{equation}
where $\bm{A}$ is a time-independent $p\times p$ square matrix, $p$ is the dimension of $\bm{X}$, and the components of $\bm{\xi}\left(t\right)$ are  white-noise, zero-mean driving terms, which satisfy the following ensemble statistics 
\begin{equation} \label{eq:EnsembleStat}
\langle \bm{\xi}_j \left(t\right) \rangle = 0, \qquad \langle \bm{\xi}_j \left(t\right) \bm{\xi}_k \left(t'\right)^{{\rm T}} \rangle =  \delta_{jk}\bm{\sigma}^2 \delta \left( t - t' \right).
\end{equation}
Here, the superscript `T' denotes the matrix transpose, and the Dirac and Kronecker delta functions are given respectively by  $\delta \left( t - t' \right)$ and $\delta_{jk}$ with $j,k = 1,\hdots, p$. The Kronecker delta function in Equation (\ref{eq:EnsembleStat}) ensures that no cross-correlations exist between the components of $\bm{\xi}\lr{t}$. This assumption is not mandatory astrophysically and is introduced in this paper to reduce the number of unknown parameters to suit the volume of X-ray timing data typically available at present. \par 
Equation (\ref{eq:generalODE}) is equivalently expressed in terms of increments of Brownian motion $\textnormal{d}\bm{B}(t)$ as an Ornstein–Uhlenbeck process \citep{Gardiner_1985} through the differential relations
\begin{equation} \label{eq:Orn_process}
    \textnormal{d} \bm{X} = \bm{A} \bm{X} \textnormal{d}t + \bm{\sigma}\;\textnormal{d}\bm{B}(t).
\end{equation}
Equation (\ref{eq:Orn_process}) exhibits a known solution; see Equation (9) in \cite{Meyers_2021}. The solution satisfies the following recursion relations
\begin{equation} \label{eq:StateSpacesoln}
    \bm{X}(t_n)= \bm{F}_{n-1} \bm{X}(t_{n-1}) + \bm{\eta}_{n-1} ,
\end{equation}
where the state transition matrix $\bm{F}_n$ and additive noise $\bm{\eta}_n$ are given explicitly by
\begin{equation}
\bm{F}_n = e^{\bm{A} \left( t_{n+1} - t_n\right)},
\end{equation}
\begin{equation}
\bm{\eta}_n = \int^{t_{n+1}}_{t_n} \textnormal{d}\bm{B}(t') \, \bm{\sigma} \, e^{\bm{A} \left( t_{n+1} - t'\right)}.
\end{equation}
\par 
In general the state variables $\bm{X}$ in X-ray timing experiments are hidden and not observed directly. Rather, they are mapped to the measurements $\bm{Y}$, viz.
\begin{equation} \label{eq:MeasurementEq}
    \bm{Y}(t_n) = \bm{C}\bm{X}(t_n) + \bm{N}(t_n),
\end{equation}
where the observation matrix is denoted by $\bm{C}$. In what follows we make the following assumptions. (i) The noise terms in Equations (\ref{eq:StateSpacesoln}) and (\ref{eq:MeasurementEq}) are normally distributed, zero-mean vectors with covariances  $\bm{Q}$ and  $\bm{\Sigma}$, i.e.\ $\bm{\eta} \sim \mathcal{N}\left( \bm{0}, \bm{Q} \right)$ and $\bm{N} \sim \mathcal{N}\left( \bm{0}, \bm{\Sigma} \right)$. (ii) The matrices $\bm{F}, \bm{Q},$ and $\bm{\Sigma}$ are functions of time $t$ and a set of static parameters $\bm{\Theta}$ to be inferred, i.e.\ $\bm{F} = \bm{F}\lr{t, \bm{\Theta}}$, $\bm{Q} = \bm{Q}\lr{t, \bm{\Theta}}$, and $\bm{\Sigma} = \bm{\Sigma}\lr{t, \bm{\Theta}}$. \par  Given the linear system of differential equations (\ref{eq:generalODE}), its associated solution (\ref{eq:StateSpacesoln}), the process and measurement noise covariance matrices $\bm{Q}$ and $\bm{\Sigma}$, together with a set of noisy observations (\ref{eq:MeasurementEq}), a linear Kalman filter can be applied to perform state estimation. In its simplest form, state tracking with a Kalman filter involves a predict and an update stage. The state and covariance are propagated using the state transition matrix according to 
\begin{equation}
\bh{X}^{-}_n = \bm{F}_{n-1} \bh{X}_{n-1},
\end{equation}
\begin{equation}
\bm{P}^{-}_n = \bm{F}_{n-1} \bm{P}_{n-1}\bm{F}^{\rm T}_{n-1} + \bm{Q}_{n-1}.
\end{equation}
The state update stage then uses the measurement at time $t_n$ to update the state and covariance, viz.
\begin{equation}
    \bh{X}_n = \bh{X}^{-}_{n} + \bm{k}_n \lr{\bm{Y}_n - \bm{C} \bh{X}^{-}_n}, 
\end{equation}
\begin{equation}
    \bm{P}_n = \lr{\bm{I} - \bm{k}_n \bm{C}} \bm{P}^{-}_n. \label{Eq:KFCovariance}
\end{equation}
The Kalman gain $\bm{k}_n = \bm{P}^{-}_n \bm{C}^{\rm T} \lr{\bm{C}\bm{P}^{-}_{n}\bm{C}^{\rm T} + \bm{\Sigma}_{n}}^{-1}$ is defined to minimize the trace of the a posteriori state error covariance \citep{Welch_1995}, which is equivalent to minimizing the squared error $|\bm{X}_n - \bh{X}_n|^2$. The derivation of the Kalman gain is provided in standard textbooks and is not repeated here; see for example \cite{Gelb_1974}.  \par 
\subsection{Parameter estimation}\label{App:ParameterEstimation}
In addition to state tracking, a further practical application of the Kalman filter is static parameter estimation.  The Kalman filter likelihood is given by \citep{Meyers_2021}
\begin{equation} \label{eq:Likelihood}
    p(\{\bm{Y}_n\}_{n=1}^{N}|\bm{\Theta}) = \prod^{N}_{n=1}\mathcal{N}(\bm{C}\bm{\hat{X}}^{-}_{n},\bm{s}_n|\bm{\Theta}),
\end{equation}
where $\bm{s}_n$ denotes the covariance of the innovation vector.  In practice we work with $\log p(\{\bm{Y}_n\}_{n=1}^{N}|\bm{\Theta})$, which is equivalent to Equation (\ref{eq:LikelihoodMain}) in the main body of the text.
\par The Kalman filter likelihood is combined with the prior distribution $p\lr{\bm{\Theta}}$  to estimate the posterior on the parameters $\bm{\Theta}$ according to Bayes' rule 
\begin{equation} \label{eq:BayeRule}
    p\lr{ \bm{\Theta}| \{\bm{Y}_n\} } = \frac{p(\{\bm{Y}_n\}|\bm{\Theta})  p\lr{\bm{\Theta}}  }{Z\lr{ \{\bm{Y}_n\} }}, 
\end{equation}
where the denominator on the right-hand side of Equation (\ref{eq:BayeRule}) is known as the evidence and is defined according to 
\begin{equation} \label{eq:evidence}
    Z \lr{  \{\bm{Y}_n\} } = \int \textnormal{d}\bm{\Theta} \, p(\{\bm{Y}_n\}|\bm{\Theta}) p\lr{\bm{\Theta}}  \; .
\end{equation} \par
The primary goal of nested sampling is to numerically approximate Equation (\ref{eq:evidence}). Posterior samples are an important by-product. The specific steps of the algorithm are sketched out in Section \ref{Sec:KFAlg}. 
\section{Alternative state-space variables}\label{App:AlternativeSSVars}
It is possible in principle to employ a different set of state-space variables to replace those introduced in Section \ref{sec:AD}, i.e.\ $\mathbf{X} = [\Omega(t), Q(t), S(t), \eta(t)]$. For example, one may elect to replace $S(t)$ with $R_{\rm m}(t)$ and Equation (\ref{eq:LE_S}) with a mean-reverting Langevin equation for $R_{\rm{m}}(t)$, viz.\ 
\begin{equation}\label{Eq:RmStochastic}
    \frac{\tn{d}R_{\rm{m}}}{\tn{d}t} = -\gamma_{R_{\rm{m}}} [R_{\rm{m}}(t) - R_{{\rm{m}}0}] + \xi_{R_{\rm{m}}}(t), 
\end{equation}
where $\gamma_{R_{\rm{m}}}$ and $\xi_{R_{\rm{m}}}(t)$ are interpreted analogously to the damping constants and fluctuating driving terms in Equations (\ref{eq:LE_Q}) and (\ref{eq:LE_S}). The Kalman filter analysis proceeds as in Section \ref{sec:AD} with the following modifications: (i) $S_0$ is replaced with $(2 \pi)^{-1} \mu^2 R_{\rm{m0}}^{-6}$; and (ii) the linearized equations (\ref{eq:EoMTrunc}) are replaced with 
\begin{equation} \label{eq:EoMTruncRmReplace}
    \frac{\tn{d}}{\tn{d}t}
    \begin{pmatrix}
    \Omega_1 \\ Q_1 \\ R_{{\rm{m}}1}
    \end{pmatrix}
    =
    \begin{pmatrix}
    -\gamma_\Omega & 0 & -3\gamma_\Omega/2 \\
    0 & -\gamma_Q & 0 \\
    0 & 0 & -\gamma_{R_{\rm{m}}}
    \end{pmatrix}
    \begin{pmatrix}
    \Omega_1 \\ Q_1 \\ R_{{\rm{m}}1}
    \end{pmatrix}
    + 
    \begin{pmatrix}
    0 \\ Q_0^{-1}\xi_Q \\ R_{\rm{m0}}^{-1}\xi_{R_{\rm{m}}}
    \end{pmatrix}.
\end{equation} \par
One subtle astrophysical implication of replacing Equation (\ref{eq:LE_S}) with Equation (\ref{Eq:RmStochastic}) is as follows. Equation (\ref{Eq:RmStochastic}) implies that fluctuations in $R_{\rm{m}}(t)$ are driven stochastically and independently from fluctuations in $Q(t)$ and $S(t)$, e.g.\ due to fluctuations as the companion star evolves or instabilities in the magnetic geometry at the disk-magnetosphere boundary, whereas Equation (\ref{eq:AlfvenRad}) implies that $R_{\rm m}(t)$ is slaved to $Q(t)$ and $S(t)$. Hence the two approaches are different physically and lead to different results in general. It is impossible to say at present, theoretically or observationally, whether real HMXBs are approximated better by Equation (\ref{eq:LE_S}) or (\ref{Eq:RmStochastic}); even three-dimensional magnetohydrodynamic simulations are not yet equipped to offer a clear-cut answer \citep{Romanova_2015}, and data volumes are too small to return a clear-cut preference from Bayesian model selection at present. \par
By selecting $\mathbf{X} = [\Omega(t), Q(t), S(t), \eta(t)]$ as the state-space variables, as in this paper, it is possible to see how $R_{\rm{m}}(t)$ and $R_{\rm{c}}(t)$ evolve with time by substituting their time-resolved histories from the Kalman filter into Equations (\ref{eq:AlfvenRad}) and (\ref{eq:CoRot}), respectively. One potential avenue for future work, once more data become available, is Bayesian model selection between the state-space framework presented in Section \ref{sec:AD} and one involving alternative state-space variables; see Section 3 in \cite{Thrane_2019} for an overview on Bayesian model selection in the context of gravitational wave astronomy. \par
We focus on the Maxwell stress in this paper because it is interesting astrophysically and notoriously difficult to measure. For example, the present analysis reveals important information about how $S(t)$ evolves in time and its relationship with (for example) the measured pulse period, e.g.\ we find moderate evidence for an anticorrelation with $r[S_1(t),P_1(t)] \approx -0.36 \pm 0.032$ in Section \ref{Sec:KFStateTracking}. Extending the present analysis to a larger population of SMC HMXBs has the potential to reveal important information about the accretion physics and $S(t)$ in particular at the disk-magnetosphere boundary.
\section{Dynamical mode of traditional Ghosh-Lamb analysis}\label{App:GhoshLambApp}
The Kalman filter analysis presented in Section \ref{Sec:KFAnalysis} is not the only way to extract time-dependent information from $P(t)$ and $L(t)$ fluctuations and hence infer $\mu$. In this appendix, we discuss one alternative studied previously by other authors, namely the dynamical mode of traditional Ghosh-Lamb analysis \citep{Yatabe_2018}. \par 
The traditional technique to infer $\mu$ using the time-averaged observables $\langle P(t) \rangle = P_0$ and $\langle L(t) \rangle = L_0$ was generalized by \cite{Yatabe_2018} to infer the magnetic field strength $B$ and mass $M$ of the HMXB X Persei. The latter authors numerically estimated a $\tn{d}P/\tn{d}t$ time series, calculated from $\tn{d}P/\tn{d}t = [P(t_{n+1}) - P(t_n)]/(t_{n+1} - t_{n})$, and inferred the underlying, static parameters by $\chi^2$ minimization. The technique, when applied to Equation (1) in \cite{Yatabe_2018}, yields a range of $B$ and $M$ estimates which return the same $\chi^2$ minimum, with $5.0 \leq B/(10^{13} \, {\rm G}) \leq 23$. \par
The dynamical Ghosh-Lamb analysis performed by \cite{Yatabe_2018} complements time-averaged techniques and Kalman filter analyses. In this appendix, for the sake of completeness and by way of comparison, we apply the dynamical Ghosh-Lamb analysis to infer $\mu$ for the X-ray transient SXP 18.3. In what follows we assume fixed canonical values for the mass $M$ and radius  $R$ of the star and generate the $\tn{d}P/\tn{d}t$ time series numerically from $P(t_n)$ measurements in the top panel of Figure \ref{fig:KFTracking}. We encourage the reader to consult Equation (1) and Section 4 in \cite{Yatabe_2018} for fuller details of the analysis.  \par
\begin{figure}
\centering{
\hspace*{-2.5cm}
    \includegraphics[width=1.3\textwidth, keepaspectratio]{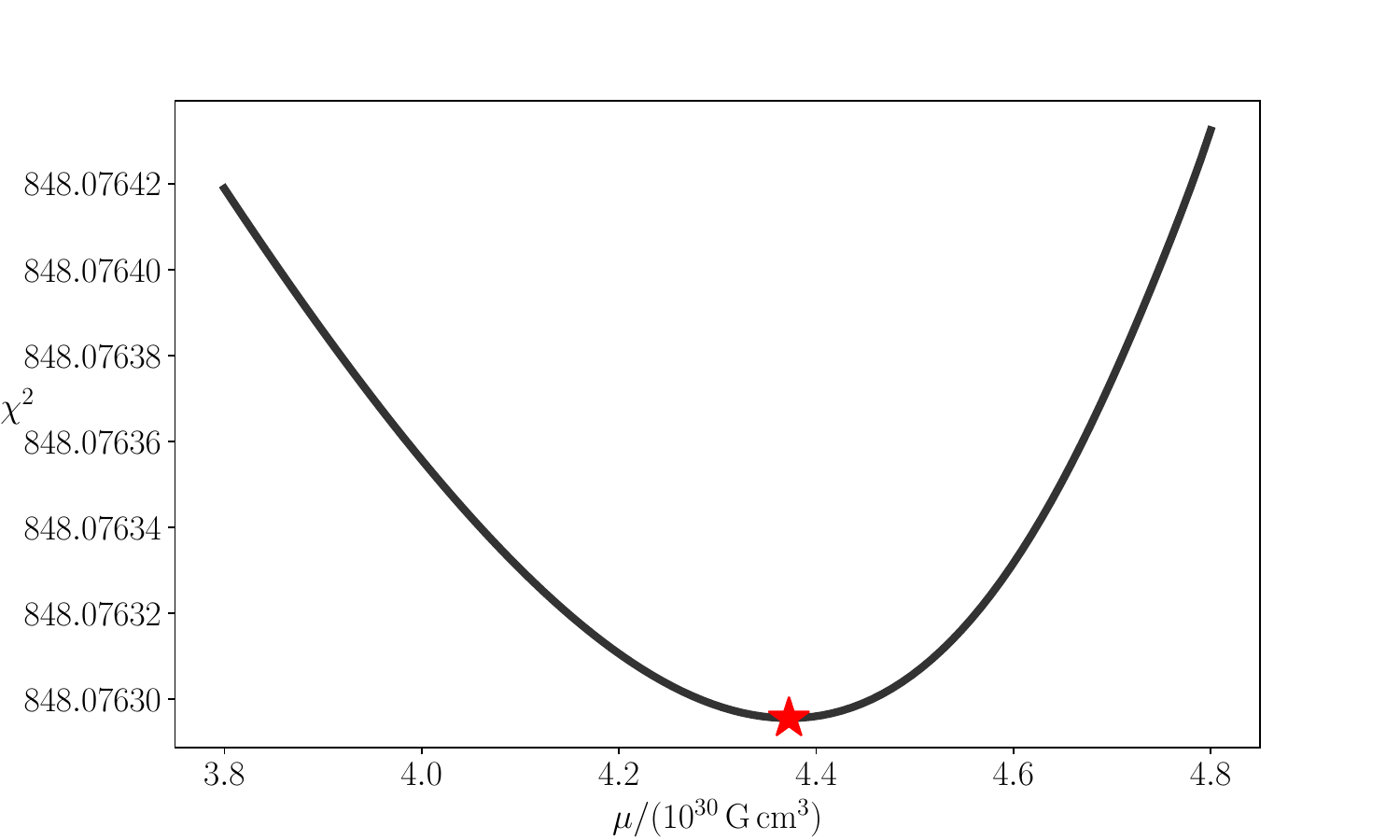}}
    \caption{Magnetic dipole moment $\mu$ estimated by $\chi^2$ minimization for SXP 18.3 following \cite{Takagi_2016} and \cite{Yatabe_2018}. The minimum (red star) occurs at $\mu \approx 4.4 \times 10^{30} \, \rm{G \, cm^3}$. }
    \label{fig:ChiSquared}
\end{figure}

 
In Figure \ref{fig:ChiSquared} we graph $\chi^2$ as a function of $\mu$. The minimum $\chi^2$ occurs at $\mu \approx 4.4 \times 10^{30} \, \rm{G \, cm^3}$, to be compared with $\mu = 8.0^{+1.3}_{-1.2} \times 10^{30} \, \rm{G \, cm^3}$ inferred in Section \ref{Sec:KFMagMom} with a Kalman filter,  and $\mu = 5.0^{+1.0}_{-1.0} \times 10^{30} \, {\rm{G \, cm^3}}$ inferred by \cite{Klus_2014} using time-averaged data and assuming $\eta_0 = 1$. It is encouraging that the three $\mu$ estimates are comparable for two reasons. (i) SXP 18.3 is classified in Table 3 in \cite{Yang_2017} as being near magnetocentrifugal equilibrium, with a pulse period standard deviation satisfying $\sigma_P = 0.025 \, {\rm s} \ll P_0$ (i.e.\ linear regime). (ii) The three estimates are derived from similar, Ghosh-Lamb-type models of magnetocentrifugal disk accretion. \cite{Klus_2014} set $\tn{d}P/\tn{d}t = 0$ to infer $\mu$; likewise, point (i) above implies $\tn{d}P/\tn{d}t \approx 0$, when minimizing $\chi^2$ in Figure \ref{fig:ChiSquared}. \par 
We do not expect the output from both techniques to coincide in general. For example, the Kalman filter features temporal autocorrelations between successive samples $Q(t_n),Q(t_{n+1}),\hdots$ [and indeed $S(t_n), S(t_{n+1}), \hdots $], caused by the factors $\gamma_Q$ and $\gamma_S$, whereas $\chi^2$ minimization does not; it treats every sample independently. Although detailed comparisons between $\chi^2$ minimization and the Kalman filter are beyond the scope of the current paper, we refer the interested reader to Sections 2--4 in \cite{Zarchan_2005} for such details. In the future, it will be interesting to compare the output from the nonlinear Kalman filter analysis in Appendix B in \cite{Melatos_2022} with the output from $\chi^2$ minimization and the dynamical Ghosh-Lamb analysis for more of the SMC HMXBs classified by \cite{Yang_2017}.
\end{document}